\documentclass[a4paper,11pt]{article}
\pdfoutput=1 
\usepackage{jcappub} 
\usepackage[T1]{fontenc} 
\usepackage{multirow}
\usepackage{orcidlink}
\usepackage{subcaption}
\usepackage{academicons}
\definecolor{orcidlogocol}{HTML}{A6CE39}
\usepackage{aas_macros}

\title{\boldmath{A Diffused Background from Axion-like Particles in the Microwave Sky}}

\author{Harsh Mehta\orcidlink{0009-0007-4664-4820},}
\author{and Suvodip Mukherjee\orcidlink{0000-0002-3373-5236}}

\affiliation{Department of Astronomy and Astrophysics, Tata Institute of Fundamental Research, Homi Bhabha Road, Mumbai, 400005, India}

\emailAdd{harsh.mehta@tifr.res.in}
\emailAdd{suvodip@tifr.res.in}

\begin{document}
\abstract{The nature of dark matter is an unsolved cosmological problem and axions are one of the weakly interacting cold dark matter candidates. Axions or ALPs (Axion-like particles) are pseudo-scalar bosons predicted by beyond-standard model theories. The weak coupling of ALPs with photons leads to the conversion of CMB photons to ALPs in the presence of a transverse magnetic field. If they have the same mass as the effective mass of a photon in a plasma, the resonant conversion would cause a polarized spectral distortion leading to temperature fluctuations with the distortion spectrum. The probability of resonant conversion depends on the properties of the cluster such as the magnetic field, electron density, and its redshift. We show that this kind of conversion can happen in numerous unresolved galaxy clusters up to high redshifts, which will lead to a diffused polarised anisotropy signal in the microwave sky. The spectrum of the signal and its shape in the angular scale will be different from the lensed CMB polarization signal. This new polarised distortion spectrum will be correlated with the distribution of clusters in the universe and hence, with the large-scale structure. The spectrum can then be probed using its spectral and spatial variation with respect to that of the CMB and various foregrounds. An SNR of $\sim$ 4.36 and $\sim$ 93.87 are possible in the CMB-S4 145 GHz band and CMB-HD 150 GHz band respectively for a photon-ALPs coupling strength of $g_{a \gamma} = 10^{-12} \, \mathrm{GeV}^{-1}$ using galaxy clusters beyond redshift $z = 1$. The same signal would lead to additional RMS fluctuations of $\sim \mathrm{7.5 \times 10^{-2} \, \mu K}$ at 145 GHz. In the absence of any signal, future CMB experiments such as Simons Observatory (SO), CMB-S4, and CMB-HD can put constraints on the coupling strength better than current bounds from particle physics experiment CERN Axion Solar Telescope (CAST).}

\maketitle
\flushbottom

\section{Introduction}
\label{sec:intro}
The Cosmic Microwave Background (CMB) is the primordial radiation that surrounds us and is a remnant of the hot Big Bang in the early universe. The initially tightly coupled radiation and baryons were decoupled due to the expansion of the universe. When the optical depth at the time of decoupling was significantly lowered, the photons were able to travel large distances and left the photon-baryon fluid at the last scattering surface at a redshift $z \approx 1089$ \cite{Dodelson:2003ft}. These photons are now observed as the CMB and are highly redshifted with a monopole temperature of 2.7255 K \cite{Fixsen_2009}. 
Many processes affect the CMB at higher multipoles like Doppler shift, lensing, scattering, etc. The power spectrum of the CMB is very well known from the correlation between temperature and polarization power spectra \cite{PhysRevD.104.022003}. Cosmic Background Explorer (COBE),  Wilkinson Microwave Anisotropy Probe (WMAP) and Planck have provided invaluable information about the CMB and our universe. Several upcoming experiments such as Simons Observatory (SO)\cite{Ade_2019}, CMB-S4\cite{abazajian2016cmbs4}, CMB-HD \cite{sehgal2019cmbhd} will be making even higher resolution CMB observations in the coming years, with an emphasis on probing the polarized CMB and spectral distortions \cite{cyr2024cmbspectraldistortionsmultimessenger,Lucca:2023seh,chluba2016spectral,chluba2013distinguishing}, which refer to the departures from Planck black-body spectrum \cite{mather1994measurement,2014PTEP.2014fB107T}.

Axions are mass possessing pseudo Nambu-Goldstone bosons that solve the strong CP problem and possess mass, making them of great interest to solve the dark matter problem \cite{berezhiani1991cosmology,dine1983not,abbott1983cosmological,preskill1983cosmology}. Their production requires high energies in particle colliders \cite{Ghosh:2023xhs}, hence probing their weak coupling with photons is a great way to detect them. In various studies, the existence of axions or ALPs will lead to spectral distortions in the CMB  \cite{marsh2014model,marsh2017axions,tashiro2013constraints}. 

The CMB photons as they pass through galaxy clusters, can undergo conversions to  ALPs if ALPs exist in the universe, irrespective of whether they constitute a fraction of dark matter \cite{Ghosh:2022rta,1992SvJNP..55.1063B,khlopov1999nonlinear,sakharov1994nonhomogeneity,sakharov1996large}. These conversions can be resonant or non-resonant and are well studied \cite{Mukherjee_2019,Mukherjee_2018,Mukherjee_2020, osti_22525054,song2024polarizationsignalsaxionphotonresonant}. The resonant conversions are dominant and require the effective masses of photon and ALP to be equal. This sets the stage for the probability of conversion that depends on the magnetic field and electron density profile, as well as the redshift of the cluster. The conversion leads to a new type of polarized spectral distortion in the CMB. With the ongoing and upcoming experiments, we will be able to gain further insight into the anisotropies and distortions in the CMB. In this paper, we study the capabilities of these experiments in being able to probe this ALP distortion signal from unresolved galaxy clusters. 

A multi-band approach can be used to constrain the ALP coupling constant from clusters that are resolved in multiple frequency observations. Radio, X-rays and optical surveys can provide information about the cluster magnetic field, electron density profiles and their redshifts respectively \cite{GOVONI_2004,Birkinshaw_1999,2014ApJS..210....9B}. The ALP distortion signal can then be probed either using the power spectrum of the region around the cluster or via a pixel signal-based approach. These will be considered in a follow-up analysis \cite{mehta2024power,Mehta:2024:new3}. The bounds on coupling constants may be independently revisited using the analysis for resolved clusters \cite{mehta2024power} or from unresolved ones, which we deal with in this work.

The unresolved clusters refer to the clusters that are not resolvable in the required electromagnetic (EM) bands (such as radio, microwave, optical, and X-ray) from which information about the electron density, magnetic field, and redshifts can be inferred. These clusters need to be resolvable also in the microwave region of the EM spectrum though, so that polarization information is measured using CMB experiments. Most of the contribution to the diffused background comes from the high redshift $z>1$ clusters for which there is a CMB measurement of the polarization signal, but no information in radio, optical, and X-ray to know the astrophysical properties and source redshift of the galaxy cluster. These signals originating from these high redshift clusters will contribute to an ALP background signal in CMB across the sky.

In this work, we show that a new kind of CMB polarised fluctuation can appear from unresolved galaxy clusters. We have studied a halo-based ALP power spectrum model is this analysis applicable to different masses of ALPs. In a halo model, the dark matter halos are biased tracers of the matter field in the universe. The galaxy clusters are embedded in these dark matter halos. These clusters are sites of photon-ALP resonant conversions.  
Thus, the large-scale structure of the universe provides a way of probing this ALP signal from unresolved clusters, using correlations between the signals at different locations in the sky. 
This correlation attributes its origin to the matter distribution in the universe. This correlation can be modeled to be within the cluster (one-halo term)  or in two different clusters (two-halo term). The one-halo term or the Poissonian component dominates at smaller angular scales, while the two-halo term or the clustering component contributes at large angular scales and may or may not dominate the Poissonian component\cite{COORAY_2002}.

The ALP background signal will depend on the ALP coupling constant, frequency of observation, ALP masses, cluster distribution in the universe, etc. 
The ALP power spectrum will also depend on the electron density and magnetic field profiles of clusters. These will be related to the masses of the clusters and their evolution at different redshifts. Thus, an understanding of the astrophysical evolution of various mass galaxy clusters at different redshifts will provide better constraints on this background spectrum, but that hasn't been considered in this analysis.

 The CMB temperature fluctuations are contaminated by foregrounds, especially from the galactic plane (like dust and synchrotron emissions). All these (including the CMB) are contaminants to the ALP-distortion signal. Based on the spatial and frequency information of the power spectrum of each component, the ALP background power spectrum can be estimated. Cleaning can improve the SNR by removing the effect of these contaminants. Not only does the signal-to-noise ratio (SNR) depend on the contamination from CMB and foregrounds, but also on the instrument beam and noise.
The impact of different cleaning techniques like template matching and  Interior Linear Combination (ILC) \cite{Eriksen_2004,ilc2008internal} in reducing the effect of foregrounds and CMB on this background signal and improving its detectability has been studied. 

The motivation behind searching for the diffused ALP signal is highlighted in Sec. \ref{sec:motive}, followed by the CMB photon to ALP resonant conversion in galaxy clusters in Sec. \ref{sec:ax_cmb_convert}.  The ALP background and its sources of variation are analyzed in Sec. \ref{sec: ALP Background}. The estimator for the ALP power spectrum and the related covariance is mentioned in Sec. \ref{sec:estimator}, followed by a spectral comparison of ALP diffused spectrum with CMB and foregrounds, and the achievable SNR for various experiments in Sec. \ref{sec:fgs impact}. The constraints obtained on ALP coupling constant using ILC are mentioned in Sec. \ref{sec:constraints}. Sec. \ref{sec: conclusion} summarizes the need and the techniques that can be used to increase the detectability of this faint background signal. The power spectrum of the ALP signal produced in a single cluster is studied in Appendix \ref{sec:alpha_l}. The variation of ALP spectrum due to contribution for very high redshifts is explained in Appendix \ref{sec:z_vary}.
 The derivation for the map-based power spectrum estimator and the bounds on coupling constant using the template matching of foregrounds is provided in Appendix \ref{sec:estim_deri} and \ref{sec:tempmatch} respectively. 
We have used natural units ($\hbar = 1, c = 1, k_B = 1$) in most places, until explicitly mentioned. We have used the cosmological parameters from Planck 2015 results \cite{2016}.

\section{Motivation} \label{sec:motive}
The ALP signal which originates from photon-ALP resonant conversion in galaxy clusters will lead to polarized distortions in the CMB at low angular scales. If these clusters are at low redshifts and resolvable in multiple EM bands, the polarization information along the line of sight can be modelled to obtain bounds on the weak coupling ALPs may be having with photons.

There will also be clusters at high redshifts which will lead to polarized ALP distortions in the CMB. These clusters may not be well resolved in multiple EM bands and will lead to a kind of diffused signal across the sky with a unique spectrum in both spectral ($\nu$) and angular frequencies or multipoles ($\ell$).
The power spectrum of this ALP background signal would dominate over the CMB power spectrum at low angular scales or high multipoles, and also increase with frequency of observation in the radio and microwave regions of the EM spectrum. This distinctive property of the polarized spectrum is a signature to detect an ALP diffused background.

To obtain bounds on the background spectrum, a comparison between the observed power spectrum of the microwave sky and that of the fiducial (no ALPs case) power spectrum is required. The background spectrum will contribute to the residual of the two spectra. Using the covariance for the observed sky, one can obtain bounds on the diffused ALP spectrum from unresolved clusters. 

\begin{figure}[h!]
     \centering
\includegraphics[height=6cm,width=15cm]{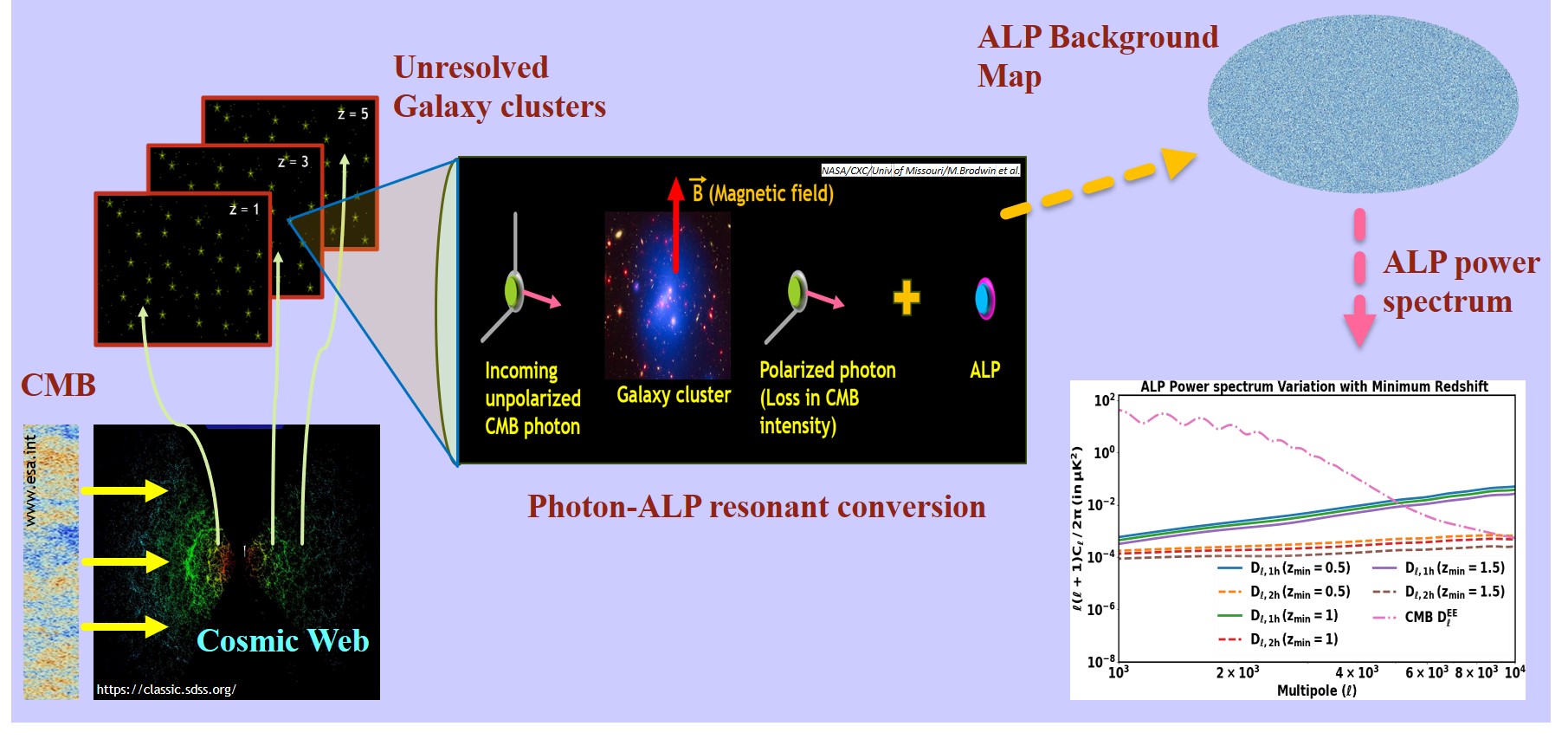}
    \caption{The following is a  schematic diagram that illustrates the production of ALP signal from unresolved clusters as the CMB photons pass through the matter field (or cosmic web) of the universe. These conversions form an ALP background signal around us, which can be modelled using the cluster distribution in our universe. The power spectrum of this ALP distortion map will be different as compared to the CMB. }
\label{fig:motivplot}
\end{figure}

The CMB photons that had scattered off the last scattering surface pass through the matter cosmic web (see Fig.\ref{fig:motivplot}). The matter power spectrum contains information about the matter density field at various scales. The matter field consists of dark matter halos, which are hosts to galaxy clusters. The clusters which are not resolved in one or some of the frequency bands or are at high redshifts, are unresolved and signals from them cannot be individually separated. The ALP distortion signal is obtained from all these clusters as the photon-ALP conversion takes place in the presence of cluster magnetic fields. These signals from unresolved clusters at various redshifts are integrated along the line of sight and produce a faint polarized ALP distortion background signal. Using halo modeling of these clusters, one can use the halo distribution to obtain constraints on the ALP background power spectrum.    
 
With the upcoming high-resolution CMB experiments such as the SO, CMB-S4, and the proposed CMB-HD, we will be able to estimate the ALP power spectrum, based on its frequency and spatial dependence, which is very distinct as compared to the spectra from other phenomena, to obtain constraints on the photon-ALP coupling constant $g_{a\gamma}$.

\section{{CMB photon - ALP resonant conversion in Galaxy clusters}} \label{sec:ax_cmb_convert}
 The CMB black-body radiation peaks at 160.2 GHz, with very tiny fluctuations. Also, it is anisotropic with spatial fluctuations in temperature of the order of $10^{-5}$. Only about $\sim 5\%$ of the CMB photons are linearly polarized. Also, there are spectral distortions in the CMB, which represent its deviations from a black-body spectrum, due to the absorption or emission of photons at different frequencies. These arise due to phenomena like the $\mu$ and $y$ distortions. Earlier the photon-baryon fluid was efficiently thermalized via processes like the Compton, double Compton and bremsstrahlung \cite{1986rpa..book.....R,mather1994measurement}. Energy release via particle-decays and primordial black hole evaporation gets redistributed. When these processes that redistribute the energy and photons, start becoming inefficient due to the Hubble expansion, distortions start setting in, which can be probed to study the early universe physics \cite{2012MNRAS.419.1294C}. The CMB polarization spectrum is mainly affected by Thomson scattering and gravitational lensing, and also exhibits an independence from the frequency of observation \cite{Hu_1997}.
 
There are also numerous secondary anisotropies that generate temperature and polarization fluctuation in the CMB. The anisotropies introduced due to galaxy clusters are secondary as they are generated after the epoch of recombination. These anisotropies can be probed either using the temperature intensity fluctuations in the CMB, or using fluctuations in its polarization. These include lensing \cite{Smith_2007}, thermal Sunyaev Zeldovich (SZ) effect \cite{Birkinshaw_1999}, kinetic SZ \cite{Birkinshaw_1999}, etc.   

In this work, we look at another secondary anisotropy, that owes its origin to galaxy clusters, and will be generated in the CMB if ALPs exist in nature. These are caused by 
the conversion of CMB photons to ALPs in the presence of the magnetic fields of galaxy clusters. These conversions are frequency-dependent and lead to spectral distortions in the CMB.
This conversion takes place in the presence of magnetic fields in astrophysical systems such as galaxy clusters and voids. We mainly focus on distortion from galaxy clusters in this analysis. The interaction between ALPs and photons is given by the following  Lagrangian \cite{Raffelt:1996wa}:

\begin{equation}\label{eq:lagr}
  \mathcal{L}_{\mathrm{int}} = -\frac{g_{a \gamma} F_{\mu \nu}\tilde{F}^{\mu \nu} a}{4} = g_{a \gamma} E \cdot B_{\mathrm{ext}} a.  
\end{equation}
This interaction introduces a polarized distortion in the CMB as an ALP is formed. After a conversion, the photon gets polarized perpendicular to the magnetic field direction at the resonant location. These conversions will be dominated by resonant conversions which satisfy the condition \cite{Mukherjee_2018}:
\begin{equation}
  m_a = m_{\gamma} = \frac{\hbar \omega_p}{c^2} \approx \frac{\hbar}{c^2}\sqrt{n_e e^2 / m_e \epsilon_{0}},
\label{fig:resonance mass}
\end{equation}
here $\omega_p$ is the plasma frequency and $n_e$ is the electron density at the location.

So, the electron density at a location in the cluster determines the ALP mass that can be formed at that location. 
For a spherically symmetric electron density profile, ALPs of a particular mass will be formed in a spherical shell in the cluster. This shell will be projected as a disk in 2D. Higher mass ALPs shall be formed near the center of the cluster with a disk of lower angular size as compared to the lower mass ALPs. The ALPs being probed are in the mass range $10^{-15} - 10^{-11}$ eV, which depends on the electron density of the cluster. 
The dispersion relation for photon-ALP conversion is given as (here $B_t$ refers to the transverse magnetic field):

\begin{equation}
\label{eq:disper}
2\omega (\omega - k) = - \omega(\Delta_e + \Delta_a) \pm \omega \Delta_{\mathrm{osc}} = \frac{m_a^2 + m_{\gamma}^2}{2} \pm \left[ \left( \frac{m_a^2 - m_{\gamma}^2}{2} \right)^2 + 
\omega^2 g_{a\gamma}^2 B_t^2  \right]^{1/2},
\end{equation}

where the parameters are defined as:
\begin{equation}
\label{eq:x}
\begin{split}
\Delta_a &= - m_a^2 / 2\omega  \,,
\qquad
\Delta_e \approx \omega_p^2 / 2\omega \,,
\\
\Delta_{a\gamma} &= g_{a\gamma}B_{t} / 2 \,,
\qquad
\Delta_{\mathrm{osc}}^2 = (\Delta_a - \Delta_e)^2 + 4\Delta_{a\gamma}^2 ,
\end{split}
\end{equation}
This dispersion relation comprises two dispersion branches.
The probability of conversion is related to the probability of a shift from one dispersion branch to the other. It is quantified using the adiabaticity parameter, which compares the scale over which electron density varies to the oscillation scale over which conversion can occur:
\begin{equation}
\gamma_{\mathrm{ad}} = \frac{\Delta_{\mathrm{osc}}^2}{|\nabla \Delta_{e}|} = \left| \frac{2g_{a\gamma }^2 B_t^2 \nu (1 + z)}{\nabla \omega_p^2} \right|. 
\label{eq:gamma_ad}
\end{equation}

If the resonant condition is satisfied ($m_a = m_{\gamma}$), in the adiabatic case ($\gamma_{\mathrm{ad}} >> 1$), the photon-ALP conversion will definitely occur. In the non-adiabatic case ($\gamma_{\mathrm{ad}} << 1$), the conversion will be probabilistic with the probability given as $1 - p$, where $p$ refers to the probability of branch shift when passing from a high density to low density region or vice-versa. This is given as \cite{Mukherjee_2018}:

\begin{equation}
p = e^{-\pi \gamma_{\mathrm{ad}}/2}.   
\label{eq:prob}
\end{equation}

There will be a probability related to the photon-ALP conversion at any location of the cluster with the mass depending on the electron density at that location. There will always be two resonances for a CMB photon while propagating in and out of the cluster due to change in electron density (low $\rightarrow$ high and then high $\rightarrow $ low ) \cite{Mukherjee_2018}. For the ultimate product of the two resonances to be an ALP, the conversion should take place at only one of the resonances. This probability is found to be \cite{Mukherjee_2020}

\begin{equation}
    P(\gamma \rightarrow a) = 2p(1 - p) = 2e^{-\pi \gamma_{\mathrm{ad}} / 2}(1 - e^{-\pi \gamma_{\mathrm{ad}} / 2}),
   \label{eq: Probab}
\end{equation}

where $1-p$ is the probability of conversion at the resonant location.

In the non-adiabatic limit ($\gamma_{\mathrm{ad}} << 1$), this is approximated as: $P(\gamma \rightarrow a) \approx \pi \gamma_{\mathrm{ad}} $.
The change in intensity due to this conversion in the CMB is given as:
\begin{equation}
    \Delta I_{\nu}^{\alpha} = P(\gamma \rightarrow a) I_{\mathrm{cmb}}(\nu) \approx \pi \gamma_{\mathrm{ad}} \left( \frac{2h \nu^3}{c^2}  \right) \frac{1}{e^{h\nu /k_BT_{\mathrm{cmb}}} - 1}.
    \label{eq:Distort}
\end{equation}

\section{Diffused Axion Power Spectrum: A new signal in the polarization sky of CMB}\label{sec: ALP Background}
The matter field in our universe comprises dark matter halos, which are hosts to galaxy clusters, which themselves are hosts to galaxies.
Any diffused signal that owes its origin to these matter overdensities can be studied using modeling of the distribution of these overdensity regions.
In the halo model, halos are descriptors of the nonlinear matter density of the universe \cite{Dodelson:2003ft,COORAY_2002,mead2015accurate}. These halos represent regions of overdensity in the matter density field, while the voids point to under-density. As we are interested in the diffused background from ALPs, we consider the halo approach to calculate the signal from high-density regions of the large-scale matter distribution in the Universe.
 
\subsection{Halo modeling of matter in the universe}

The matter power spectrum contains information about the 3D distribution of matter. It defines the matter density field at different scales. It is well described by a linear theory at large scales, while higher-order non-linear statistics are required to describe the gravitational collapsed systems at small scales, which calls for a halo modeling of the matter field in the universe. We have used the $\mathrm{\Lambda CDM}$ cosmological model to calculate the matter power spectrum using CAMB \cite{2011ascl.soft02026L}. The perturbations in the matter density field can be written as 
\begin{equation}
\rho_{m}(x) = [1 + \delta_m(x)] \bar{\rho_m},
\label{eq:denspert}
\end{equation}
with $\bar{\rho_m}$ being the mean matter density. The dark matter is the dominant constituent of the matter density in the universe. The distribution of these halos can be assumed to follow the matter distribution in the universe. This gives the matter density field from a superposition of halos $'i'$ of masses $M_i$(   \cite{Dodelson:2003ft}): 

\begin{equation}
\rho_{m}(x) = \sum_{i}\rho_h(|x - x_i|,M_i),
\label{eq:halosuper}
\end{equation}
with $\rho_h$ being the density profile of the halo.
 These halos are assumed to only be interacting gravitationally and hence, their properties only depend on their masses. These are assumed to have undergone spherical collapse and subsequent virialization. Their masses are typically defined as the mass within the radius at which the density of the halo becomes about 200 times the critical density of the universe $\rho_{c}$. The profiles start steepening beyond this radius.

The perturbations in the matter density field can be expressed in the halo model as:

\begin{equation}
\delta_m (x) = \int \mathrm{d} \, \ln \, M \frac{M}{\rho_m} \frac{\mathrm{d}n}{\mathrm{d} \,\ln \, M} \int \mathrm{d}^3 x' \delta_h(x',M)y(|x - x'|,M),
\label{eq:perturb}
\end{equation}
where $y(x,M) = \rho_{h}(x,M) / M$ is the normalized profile and  $\delta_h$ accounts for the variations in the mass function, i.e.,
\begin{equation}
\frac{\mathrm{d}n(x)}{\mathrm{d}M} = [1 + \delta_h(x,M)]\frac{\mathrm{d}n}{\mathrm{d}M}.
\label{eq:dndmpert}
\end{equation}

The distribution of dark matter halos with halo mass and redshift is referred to as the halo mass function. A general mass function is of the form 
\begin{equation}
\mathrm{d}n/\mathrm{d}M = f(\sigma)\frac{\rho_c \Omega_m}{M}\frac{\mathrm{d} \,\ln \sigma^{-1}}{\mathrm{d}M}.
\label{eq:dndm}
\end{equation}
Here $\Omega_m$ is the matter density parameter. The  $\sigma$ represents the RMS deviation in the initial density fluctuation field smoothed with a tophat filter, and $f(\sigma)$ is the halo multiplicity function. The Tinker mass function \cite{Tinker_2008} uses the multiplicity function with four free parameters ($d = 1.97,e=1.00, f =0.51,g = 1.228$) and a normalization ($B = 0.482$) and the values are set to those for $M_{200m}$. The multiplicity function reads 
\begin{equation}
f(\sigma) = B\left[ \left(\frac{\sigma}{e}  \right)^{-d} + \sigma^{-f}  \right] \exp(-g/\sigma^2) .
\label{eq:multiplicity}
\end{equation}
In the halo model, the non-linear matter power spectrum can be written as the sum of the contributions from one and two-halo terms, i.e.,
\begin{equation}
P(k) = P_{1h}(k) + P_{2h}(k).
\label{eq:netpowspec}
\end{equation}
The mass elements in a single halo are accounted for by the one-halo term, while the clustering information is contained in the two-halo term. The two-halo term dominates at low angular scales, while the one-halo term may dominate at high angular scales.

The correlation function quantifies the excess probability of finding two halos separated by some distance with respect to the Poissonian probability in the case of random uniform distribution. It is given as \cite{COORAY_2002}:
\begin{equation}
\xi_{mm}(| \textbf{x}_1 - \textbf{x}_2|) = \langle \delta_m( \textbf{x}_1)\delta_m(\textbf{x}_2) \rangle.
\label{eq:correl}
\end{equation}

The spatial matter-matter correlation function  is given as the Fourier transform of the power spectrum, i.e.,
\begin{equation}
\xi_{mm}(r) = \frac{1}{2\pi^2} \int \mathrm{d}k \, k^2 P(k) \frac{\sin(kr)}{kr}.
\label{eq:matcorr}
\end{equation}

The galaxy clusters are embedded in these dark matter halos. The distribution of these halos traces that of matter. This distribution is related using the linear bias 
(adopted from \cite{Tinker_2008}). connecting the halo-matter and matter-matter correlation functions:
\begin{equation}
\xi_{hm}(z,r) = b(z) \xi_{mm}(z,r). 
\label{eq:halocorr}
\end{equation}
\subsection{Halo Modelling of photon-ALP Resonant conversion Power Spectrum}

The CMB photon-ALP resonant conversion for unresolved clusters can be modelled with the distribution of galaxy clusters at high redshifts. The clusters can be considered to be halos of low masses ($\mathrm{10^{13} - 7 \times 10^{15} \, M_{\odot}}$), modulus a cluster bias factor taking into account the astrophysics of these clusters. We use the matter power spectrum from CAMB and the Tinker mass function for our analysis.

The ALP signal at a location within the cluster will be correlated with the signal for locations within the cluster (due to the finite probability of resonant conversion). Similar to the case of SZ power spectrum (see \cite{Komatsu_1999}), the one-halo term represents the Poissonian component of the power spectrum and is given as:

\begin{equation}
C_{\ell,1h}^{\mathrm{ax}} = \int_{z_{\mathrm{min}}}^{z_{\mathrm{max}}} \mathrm{d}z \frac{\mathrm{d}V_c}{\mathrm{d}z} \int_{M_{\mathrm{min}}}^{M_{\mathrm{max}}} \mathrm{d}M \frac{\mathrm{d}n(M,z)}{\mathrm{d}M} | \alpha_{\ell} (M,z)| ^2 .
\label{eq:onehalo}
\end{equation}
Here $\frac{\mathrm{d}V_c}{\mathrm{d}z}$ is the differential comoving volume and $\alpha_{\ell}$ is the angular harmonic transform of fluctuations (refer to Appendix \ref{sec:estim_deri}) due to the photon-ALP resonant conversion in a single cluster \cite{mehta2024power, Mukherjee_2020}. $\alpha_{\ell}$'s depend on the electron density and magnetic field profiles of the clusters.  The inference of  $\alpha_{\ell}$'s and its dependency on frequency, coupling constant, and cluster profiles is explained in Appendix \ref{sec:alpha_l} when ALPs of all masses in the range $10^{-15} - 10^{-11} $ eV are assumed to be generated in galaxy clusters. The integrals take into account the total number of clusters of various masses at different redshifts which lie in the cluster mass range ($M_{\mathrm{min}}$ to $M_{\mathrm{max}}$) and the redshift range being considered ($z_{\mathrm{min}}$ to $z_{\mathrm{max}}$). This contribution to the ALP power spectrum will be present even if there is no clustering effect due to gravitational attraction.

The ALP signal will also be correlated with the signal at locations outside its cluster. This is taken into account by the two-halo term (where we use the limber approximation applicable at $\ell > 20$) given as: 
\begin{equation}
C_{\ell,2h}^{\mathrm{ax}} = \int_{z_{\mathrm{min}}}^{z_{\mathrm{max}}} \mathrm{d}z \frac{\mathrm{d}V_c}{\mathrm{d}z} P_m \left( k = \frac{\ell +1/2}{r(z)},z \right) \times \left[  \int_{M_{\mathrm{min}}}^{M_{\mathrm{max}}} \mathrm{d}M \frac{\mathrm{d}n(M,z)}{\mathrm{d}M} b(M,z) \alpha_{\ell} (M,z) \right]^2.
\label{eq:twohalo}
\end{equation}
Here $r(z)$ is the comoving distance at redshift $z$.
The two-point halo correlation function has been expressed in terms of the matter power spectrum as
\begin{equation}
P_h(k,M_1,M_2,z) = b(M_1,z)b(M_2,z)P_m(k,z),
\label{eq:biaspow}
\end{equation}
which arises due to the  clustering between the halos.

There will be several clusters in a  certain mass interval within a given redshift interval. 
We have considered random electron density and magnetic field profiles for the galaxy clusters and the median of their $|\alpha_{\ell} (M,z)|^2$'s as the contribution of these clusters to the ALP background spectrum. In principle, these single cluster power spectra ($|\alpha_{\ell} (M,z)|^2$'s) depend on the evolution of clusters based on their masses and redshifts. This in turn would allow one to relate the cluster electron density and magnetic field profiles with their masses at different redshifts and model the single cluster contribution to the power spectrum well. The single cluster power spectra ($|\alpha_{\ell} (M,z)|^2$'s) will also depend on the range of ALP masses that are being considered to be produced in the cluster and the variation of coupling constants with the ALP masses. 
We consider the  case where
ALPs of all masses in the range $10^{-15} - 10^{-11}$ eV are produced in the galaxy clusters with the corresponding probabilities if the resonant condition is satisfied ($m_a = m_{\gamma} $). The coupling constant has been assumed to be uniform for all ALP masses in this range ($g_{a\gamma} = 10^{-12} \, \mathrm{GeV^{-1}}$). 

We have considered the ALP background signal to have its origin in clusters of masses $\mathrm{10^{13} - 7 \times 10^{15} \, M_{\odot}}$. We create various mass bins in this range. We simulate various mass binned clusters in redshift bins from $z = 0.5$ to $z = 7$, beyond which we believe the background signal won't be affected much by higher redshift clusters. We simulate the quantity $\textbf{|B|}^2 / \nabla n_e$ in clusters and select the median values at various distances from the cluster center to obtain an ALP signal that would serve as a representative for the particular bin. The power spectrum for such a cluster acts as the ALP distortion spectrum $\mathrm{|\alpha_{\ell}|^2}$ for that bin and is used in the evaluation of the one and two-halo ALP power spectra in Equs.\ref{eq:onehalo} and \ref{eq:twohalo} respectively. 

The one-halo term dominates at high multipoles as it considers the cluster interior where the signal is generated, which corresponds to smaller angular scales. 
So, there are power contributions at high redshifts from various clusters. This prevents the lowering of one-halo power spectrum as it scales as $\ell^{0}$ at high multipoles.
\begin{figure}[h!]
     \centering
\includegraphics[height=8cm,width=12cm]{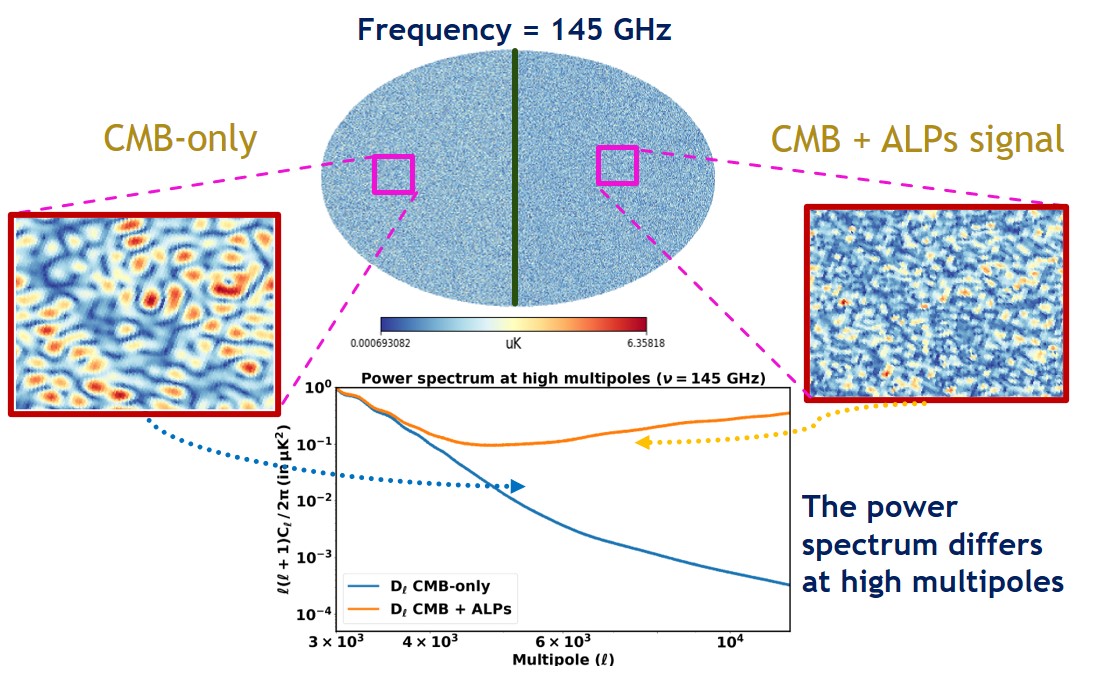}
    \caption{The CMB is smooth at low angular scales, but if photon-ALP resonant conversion takes place, it will lead to additional fluctuations in the CMB. The RMS fluctuations from this conversion for a  coupling constant of $g_{a\gamma} = 10^{-12} \, \mathrm{GeV^{-1}}$, would be  of the order of $\mathrm{7.5 \times 10^{-2} \, \mu K}$. The observed power spectrum at high multipoles in the presence of ALPs will not follow the usual dampening as observed for the case of the CMB-only map (blue line), but will increase with multipoles (orange line). }
       \label{fig:alpmap}
\end{figure}

The two-halo term increases at low multipoles ($\ell$ = 20 to 100) as the correlation is high, accompanied by a large number of clusters at low redshifts. With increasing multipoles, it then decreases as the number of clusters starts decreasing significantly with redshift and the halo correlation also decreases. The two-halo term may contribute more than one-halo term for the low-multipole range ($\mathrm{20 < \ell < 200}$) when $M_{\mathrm{min}}$ is low, as the low masses at low redshifts contribute significantly to the halo-halo correlation. 
Hence, the ALP spectrum shape will depend on the strength of the signal at low multipoles, which will be determined by the number of clusters contributing and their individual contributions, which themselves will depend on the cluster masses and profiles. 

{
There will be clusters at very high redshifts and with the polarized photon travelling through high and low redshift clusters, it may get depolarized due to turbulence or stochastic effects.  Hence, it is mostly the low redshift clusters ($z < 3.5$), (accompanied by the fact that they are in large numbers) that mainly contribute to the polarized ALP diffused signal. The polarization information will be lost as well if the beam size of the instrument is higher than the angular scale of the cluster in the sky.}

The CMB is very smooth at very high multipoles ($\ell > 4000$), with low spatial fluctuations. If there is photon-ALP conversion in unresolved clusters, it will create a diffused background of the ALP signal which will cause additional fluctuations in the CMB at low angular scales or high multipoles (see Fig.\ref{fig:alpmap}).  For a coupling constant of $g_{a\gamma} = 10^{-12} \, \mathrm{GeV^{-1}}$, the additional RMS fluctuations would be of the order of $\mathrm{7.5 \times 10^{-2} \, \mu K}$ at 145 GHz. The effect on the observed power spectrum is seen at high multipoles ($\ell > 4000$) where the CMB-only map shows a dampening of power with increase in multipoles, while the power increases in the case of the CMB $+$ ALPs map. 
\begin{figure}[h!]
     \centering
\includegraphics[height=6.5cm,width=12.5cm]{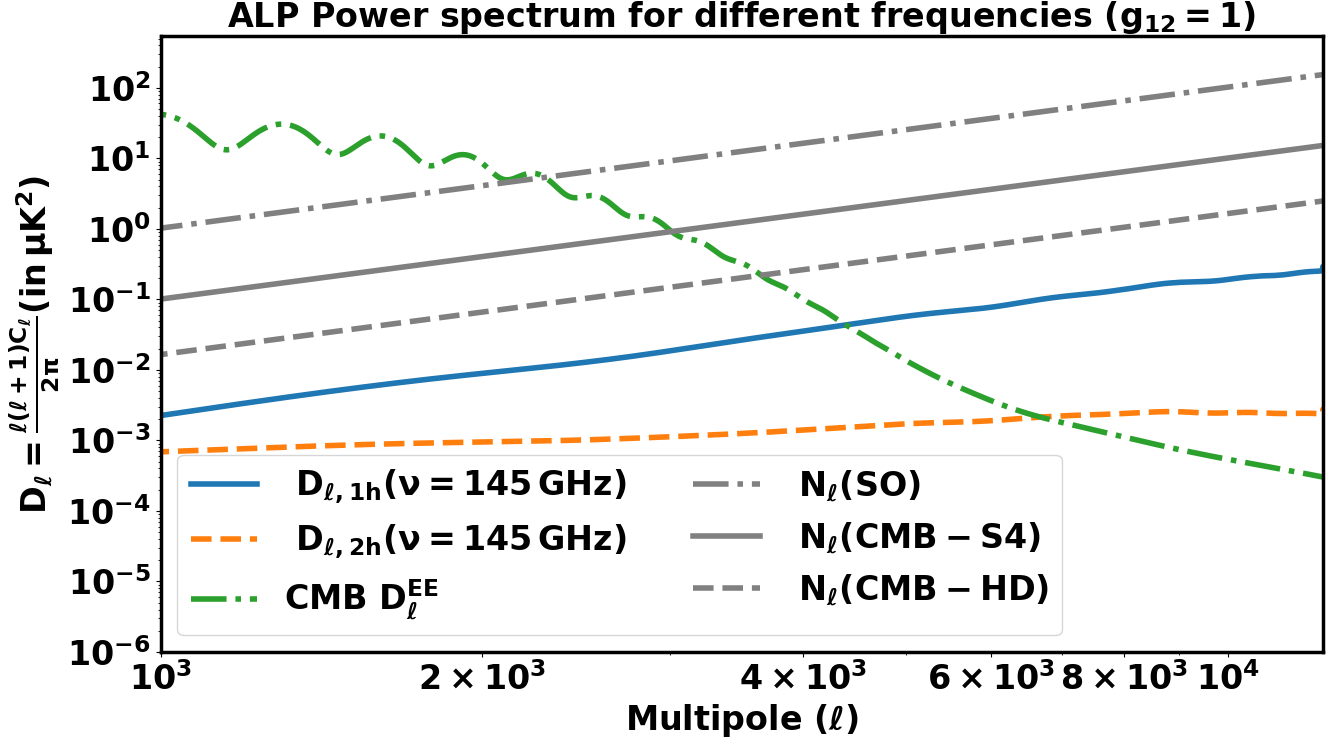}
    \caption{This figure shows the one-halo Poissonian (solid lines) and two-halo clustering (dashed lines) components of the ALP background spectrum for a coupling constant of $g_{a\gamma} = 10^{-12} \, \mathrm{GeV^{-1}}$ at 145 GHz.
    The CMB polarization spectrum $D_{\ell}^{EE}$ is shown in a dash-dotted line and is independent of frequency. Here $g_{12}$ is dimensionless defined as  $g_{12} = g_{a\gamma} \times  10^{12} \, \mathrm{GeV} $. The grey lines represent the noise power spectrum $N_{\ell}$ corresponding to the beam size for frequency bands in the range 140-150 GHz for various detectors.}
       \label{fig:f_halo_vary}
\end{figure}

The one-halo and two-halo power spectra at 145 GHz are shown in Fig.\ref{fig:f_halo_vary}.
These spectra correspond to a minimum redshift $z_{\mathrm{min}} = 1$. The grey lines represent the noise power spectrum $N_{\ell}$ corresponding to the 140 - 150 GHz band for various detectors (SO, CMB-S4, and CMB-HD) as these are the bands with higher sensitivities compared to the ones at higher and lower frequencies.

The increase in one-halo at low multipoles sets the maximum angular scale up to which ALPs are formed. 
At high multipoles, the shape of the ALP power spectrum will be independent of the strength and scale almost as $\ell^0$ ($D_{\ell}$ varies as $\ell^{2}$) following the one-halo contribution. Here we have considered a randomly uniform orientation of the magnetic field at various locations in a cluster. This gives the $\ell^{2}$ dependence to $D_{\ell}^{\alpha \alpha}$ at high multipoles.

In principle, the variation of individual $|\alpha_{\ell}|$'s at high multipoles will depend on the magnetic field orientation within the cluster. The individual variations due to these $|\alpha_{\ell}|$'s will be suppressed as many clusters are integrated along the line of sight. {The polarization information for a photon travelling from high redshift clusters may get lost due to depolarization from turbulence and stochastic effects in galaxy clusters. Thus, the low redshift clusters will impact the shape of the background spectrum the most as these are the clusters for which the polarization signal can be resolved.} The ALP background spectrum will be correlated with the synchrotron background spectrum from galaxy clusters. In this analysis, we have also considered that the coherent length scale of magnetic fields is larger than the beam size of the CMB experiments for it to be detectable. For coherent scales of magnetic field that are smaller than the beam size, the polarised signal will be smeared out. We have taken this effect into account in our analysis. Thus a high-resolution beam is important, so that the polarization information is well preserved. The polarized distortion information from high redshift clusters will be lost as the angular scale of the magnetic field will be smaller than the beam size. Thus for the upcoming CMB experiments, it is mostly the low redshift clusters that will significantly contribute to the background spectrum. With high angular resolution CMB experiments in the future, measurement of the signal from the high redshift clusters will improve by using multiple frequency channels. Efficient foreground cleaning techniques \cite{Remazeilles_2021,vacher2023high,vacher2022moment} on the polarization maps will be useful in reconstructing this signal from data in the future. In addition to the smearing due to the instrument beam, there can be smearing along the line of sight if there are multiple galaxy clusters along the same line of sight where resonant photon-ALPs conversion can take place. However, this effect is low as the probability of multiple galaxy clusters along the same line of sight is less. Also, the amplitude of the diffused spectrum will increase with the strength of the magnetic fields for these low redshift clusters.

\subsection{The differences of the ALP power spectrum from other polarized cosmological signals.}
The lensed CMB polarization power spectrum $D_{\ell}^{EE}$ is independent of the frequency of observation. Also, its dependence on multipoles is well-known from the correlation between temperature and polarization spectra. Other effects that induce polarization in the CMB include reionization and the polarized SZ effect.  The reionization power spectrum decreases with an increase in frequency and also weakens at high multipoles. The polarized SZ power spectrum increases with multipoles but its strength is weak with fluctuations of the order of $10^{-8}$ K \cite{Birkinshaw_1999, carlstrom2002cosmology}. 
The ALP power spectrum strength on the other
hand increases with frequency in the radio and microwave regimes of the electromagnetic spectrum. The two-halo term may dominate at low multipoles ($20 < \ell < 200 $), while the one-halo term dominates at high multipoles. The high multipoles can be used to probe the ALP signal from the damped CMB power spectrum.
The spectrum takes into account the ALP signals generated in various mass clusters, integrated over different redshifts. The individual features of the signal are thus suppressed and the power spectrum does not show spikes or high oscillations. This is in contrast to the CMB at low multipoles. For a low $z_{\mathrm{min}}$, this characteristic may not hold as resolved, and well-luminous clusters may be contributing to the background.    
For the mass range $10^{13}$ to $\mathrm{7 \times 10^{15} \, M_{\odot}}$, redshift range 1 to 7, and a coupling constant $g_{a\gamma} = 10^{-12} \, \mathrm{GeV^{-1}}$, the one and two-halo contributions are shown in Fig. \ref{fig:f_halo_vary}. Here $g_{12}$ is dimensionless and is defined as $g_{12} \equiv g_{a\gamma} \times 10^{12} \, \mathrm{GeV}$.
The ALP power spectrum crosses the CMB power spectrum at a lower multipole value for higher frequencies and vice versa. 
These spatial and spectral variations with respect to other polarized signals can be used to probe the ALP background spectrum. 
Also, the diffused ALP signal can potentially impact the study of small-scale B-mode polarization signals, which are mainly attributed to gravitational lensing of CMB photons due to large-scale structure \cite{Stompor_1999,aghanim2020planck}. However, as the spectral shape and the angular power spectrum due to the ALP distortion are different from CMB B-mode polarization, it is possible to distinguish the two using multiple frequency bands. Furthermore, a joint estimation of the two signals will be possible as well from the future detectors.

\subsection{Sources of variation of ALP diffused spectrum due to cosmological factors}
The background spectrum owes its origin to the unresolved clusters of various masses at different redshifts. Also, the coupling and masses of ALPs will determine the amplitude of this diffused spectrum. 

\begin{figure}[h!]
     \centering
\includegraphics[height=6.5cm,width=12.5cm]{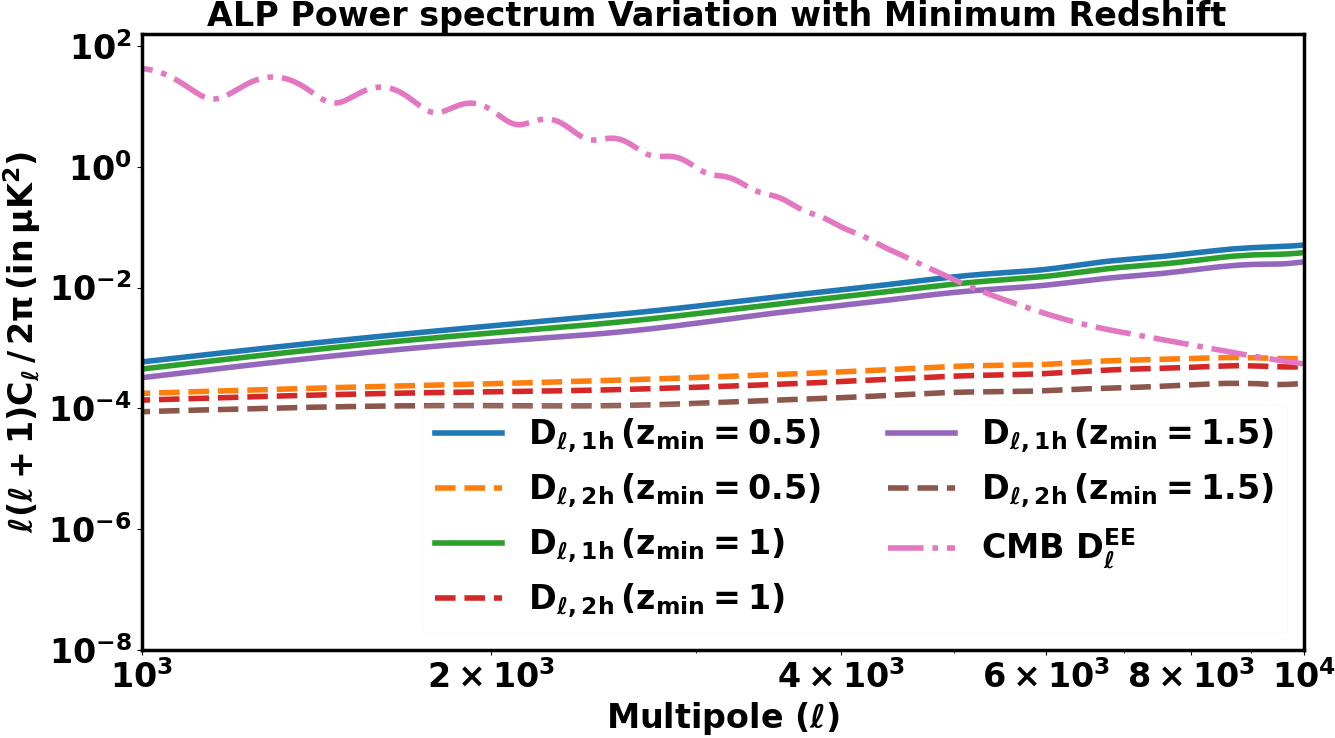}
    \caption{Variation of the ALP power spectrum (solid: one-halo, dashed: two-halo) for various minimum redshifts $z_{\mathrm{min}}$. The power spectrum strength decreases with an increase in the minimum redshift which is considered as the clusters at low redshifts contribute to the background spectrum. The CMB E-mode spectrum is also plotted for reference  }
       \label{fig:zmin_halo_vary}
\end{figure}

\textbf{Variation with minimum redshift:} With a lower minimum redshift ($z_{\mathrm{min}}$), the background power spectrum increases (Fig.\ref{fig:zmin_halo_vary}), while it decreases with a higher minimum redshift. It is mainly the low redshift clusters that contribute to the ALP background signal as these are large in number and have their polarization signals intact. For the high redshift clusters, depolarization of the photons may occur as a result of multiple scatterings. With the upcoming improved detectors, we will be able to resolve clusters at lower redshifts ($z < 1$). The bright clusters at low redshifts could also significantly affect the shape and strength of the power spectrum, especially at low multipoles. Their contribution to the background spectrum need not be considered and constraints from them can be obtained using the analysis for resolved clusters (explained in a separate work \cite{mehta2024power}). 

\begin{figure}[h!]
     \centering
\includegraphics[height=6cm,width=12cm]{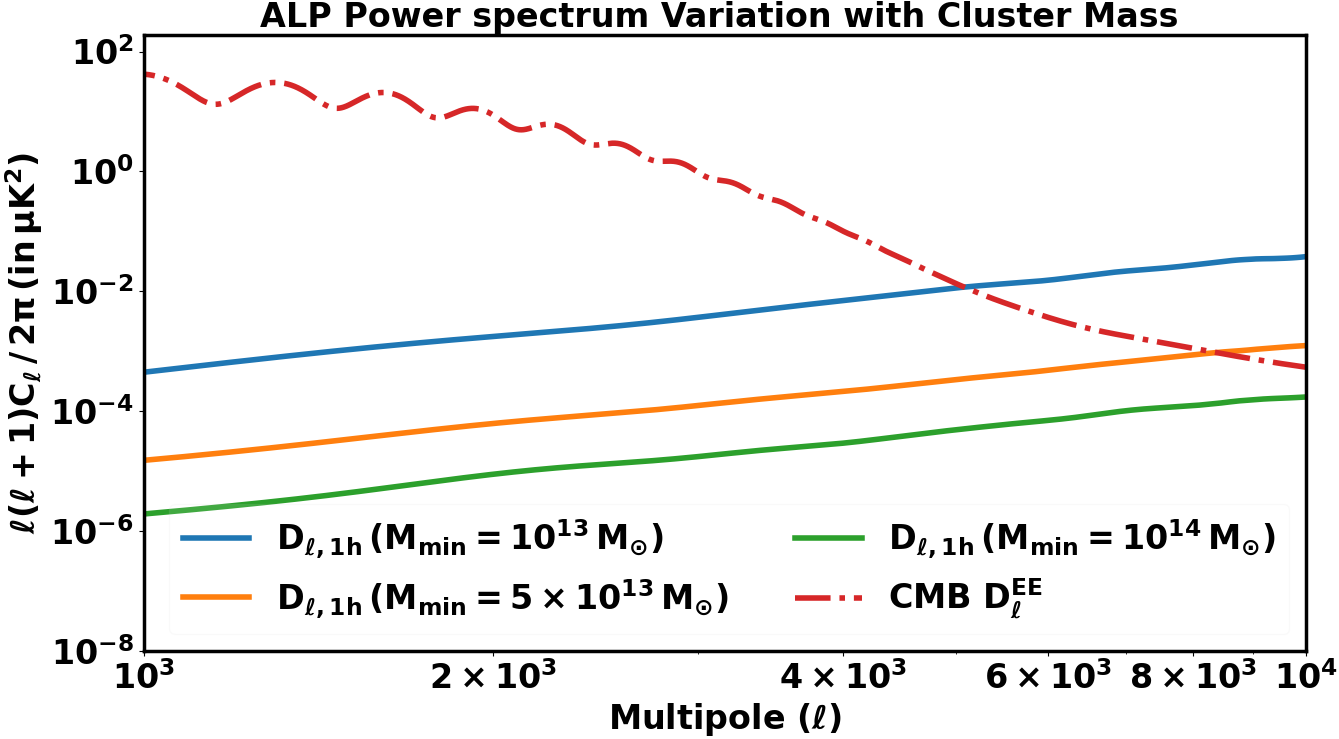}
    \caption{This figure shows the variation of the ALP background spectrum with the minimum mass considered for the cluster range ($M_{\mathrm{min}}$). The spectrum increases with a lower minimum mass. This is due to the presence of more low mass clusters as is depicted by the halo mass function. These low mass clusters contribute significantly to the spectrum, because of our random modelling of galaxy cluster profiles. }
       \label{fig:m_halo_vary}
\end{figure}
\textbf{Variation with cluster masses:}
The strength of the power spectrum also depends on the range of masses being considered. 
For different halo mass ranges, the spectrum decreases with decreasing mass range (Fig.\ref{fig:m_halo_vary}). The higher cluster masses contribute to low redshifts. The low cluster masses are less resolved and may contribute even at higher redshifts. The decrease in background spectrum will depend on the contribution of various mass clusters to the ALP background spectrum, which can be analyzed by studying the relation between cluster masses and their electron densities. Since the one-halo power spectrum is about two orders in magnitude higher than the two-halo contribution at the relevant scales corresponding to high multipoles ($\ell$ > 4000), we will not be plotting the two-halo contributions in any of the upcoming plots. But its contribution to the calculations will be considered.

\begin{figure}[h!]
     \centering
\includegraphics[height=6.5cm,width=12.5cm]{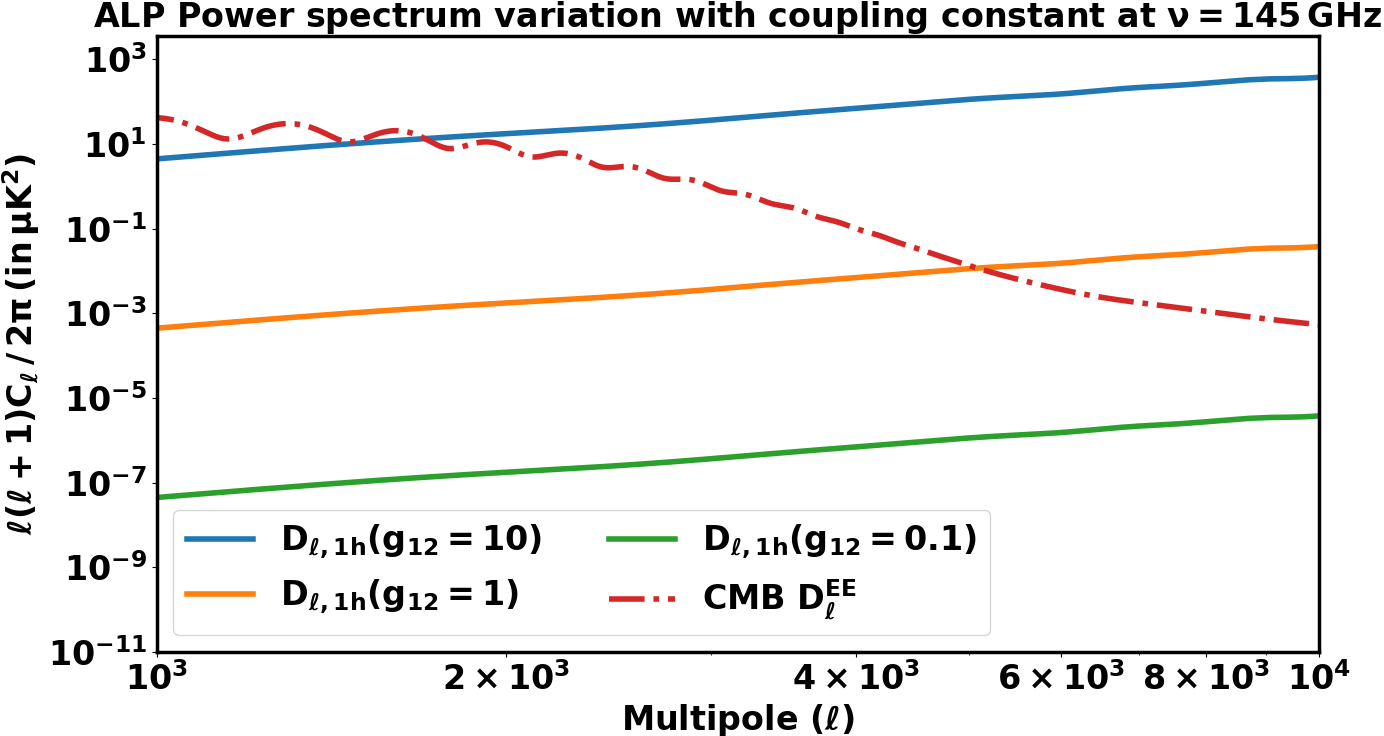}
    \caption{This figure shows the variation of ALP background spectrum  with the ALP coupling constant $g_{a\gamma}$. The power spectrum increases as the coupling strength increases, scaling as $g_{a\gamma}^4$ The CMB E-mode spectrum is also plotted for reference. Here $g_{12} \equiv g_{a\gamma} \times  10^{12} \, \mathrm{GeV} $.}
       \label{fig:ga_halo_vary}
\end{figure}

\textbf{Variation with photon-ALP coupling constant and ALPs masses:} The amplitude of the ALP signal is proportional to the square of the coupling constant  $g_{a\gamma}$ (Eq.\ref{eq:gamma_ad}), so the power spectrum scales as $g^4_{a\gamma}$ (see Fig. \ref{fig:ga_halo_vary}). This is expected as the higher the coupling, the higher will be the distortion signal. 

We have considered ALPs of all masses in the range $\mathrm{10^{-15} - 10^{-11} \, eV}$ are being produced.
The ALP background spectrum will also depend on the ALP masses that may exist in nature. If ALPs of masses only in a certain subrange of the mass range considered, exist in nature, the power spectrum shall decrease. This decrease will be greater if only high mass ALPs exist ($\mathrm{10^{-13} - 10^{-11} \, eV}$), while lower if low mass ALPs exist ($\mathrm{10^{-15} - 10^{-13} \, eV}$). This is because the single cluster distortion spectra $\alpha_{\ell}$'s
are generally higher for low mass ALPs which are formed in the outer regions of clusters with low electron density and high conversion probabilities (see Eq.\ref{eq:gamma_ad}).

If ALPs of only a particular mass exist, the resonant conversion can occur only at two locations along a line of sight. However, if ALPs of different masses co-exist, the depolarization of the polarization components of the signal can occur when the magnetic field orientation varies, and the different mass ALPs are formed along the same line of sight. This is the case for the overlapping region of the different mass ALP signal disks and lines of sight near the cluster center. But the intensity of the signal will increase in the overlapping region.

For a variation in coupling constant with ALP mass, signals would be generated in spherical shells in a spherically symmetric cluster. These shells would be visible as different-sized disks, with a larger disk corresponding to low mass ALPs. 
The intensities from these different disks will then need to be summed up to obtain the net ALP distortion signal.  

\subsection{Sources of variation in the ALP Power Spectrum due to cluster astrophysics}\label{sec:ne_B_halo}

The ALP background spectrum will depend on the single cluster power spectra ($|\alpha_{\ell}|$'s), which depend on the astrophysics of the galaxy clusters via their electron density and magnetic field profiles.
For all clusters considered, the profile parameters have been assigned random values within the allowed range. In principle, these would depend on the masses of clusters and their evolution at different redshifts. 

\begin{figure}[h!]
     \centering
\includegraphics[height=12cm,width=13cm]{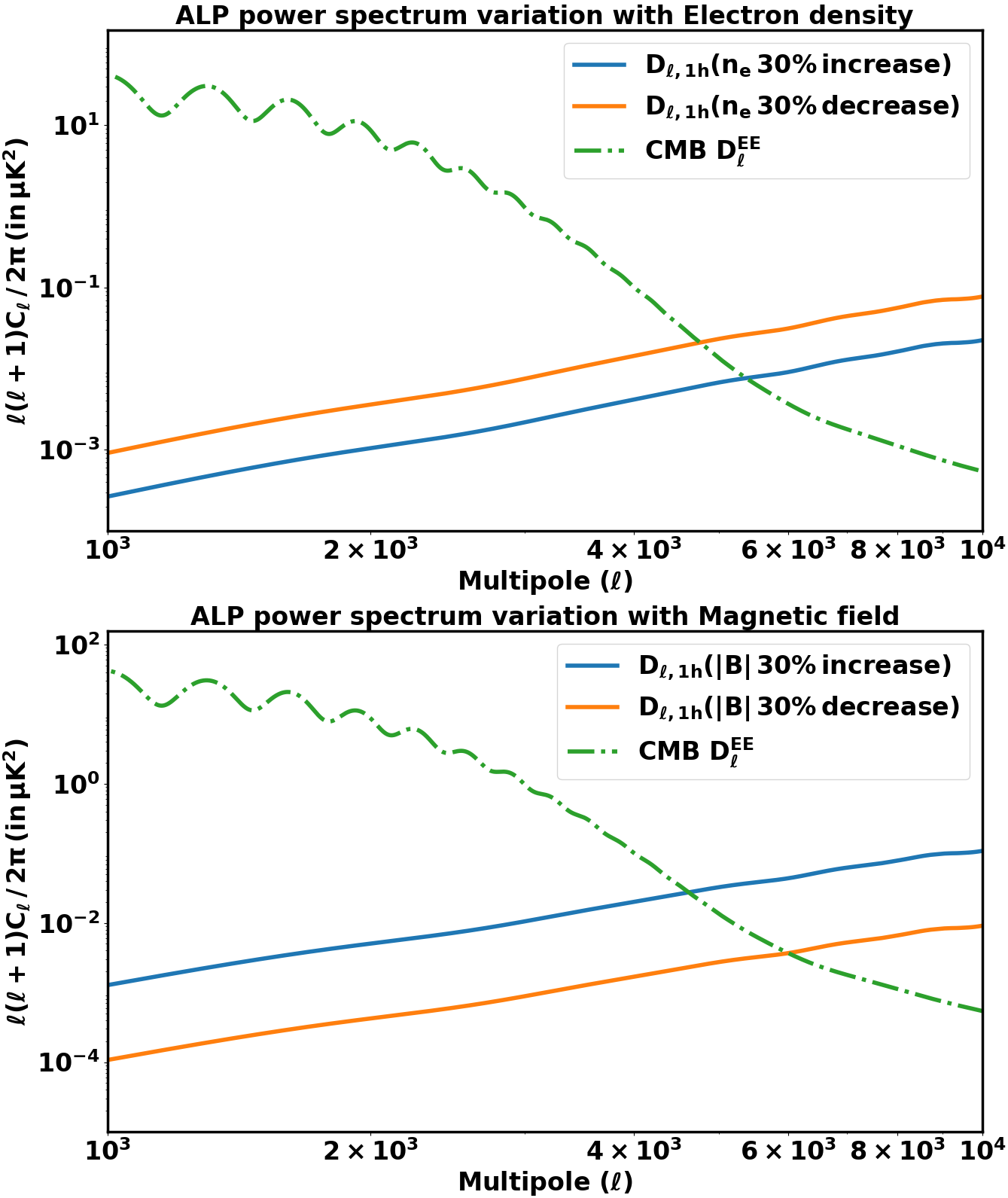}
    \caption{(a) Variation of the ALP power spectrum with electron density strength  $n_{e}$. The power spectrum strength varies inversely with the square of electron density ($D_{\ell}^{\alpha \alpha} \propto n_e^{-2}$), when all mass ALPs are formed.(b) Variation of the ALP power spectrum with magnetic field magnitude  $\textbf{|B|}$. The power spectrum strength varies proportionally with the quadruple of magnetic field ($D_{\ell}^{\alpha \alpha} \propto \mathrm{\textbf{|B|}^{4}}$). The CMB E-mode spectrum is also plotted for reference.  }
       \label{fig:neB_halo_vary}
\end{figure}

The power spectrum depends on the electron density profiles, magnetic field profiles,  the masses of galaxy clusters, and the dependence of electron density on the mass density of galaxy clusters. The effect of change in profiles will be similar to the dependence followed by $|\alpha_{\ell}|^2$, as expected. 
If ALPs of masses in only a sub-range of what we have considered ($10^{-15} - 10^{-11}$ eV) are formed, the dependence of the ALP background spectrum on these profile parameters can change, especially in the case of high mass ALPs, following the variation of $|\alpha_{\ell}|^2$. The variation of $|\alpha_{\ell}|^2$  with the mass of ALPs has been considered in a separate work \cite{mehta2024power}.

Here we consider the variation of ALP diffused spectrum with a change in electron density and magnetic field strengths. With the mass range of $10^{-15} - 10^{-11}$ eV, ALPs can be formed at every location for most of the galaxy clusters. Since the probability of conversion goes as $P(\gamma \rightarrow a) \propto |\nabla n_e|^{-1}$, the signal strength at every location within the cluster will vary as $\propto n_e^{-1}$. Thus the background power spectrum shall vary as $D_{\ell} \propto {n_e^{-2}}$ (see Fig. \ref{fig:neB_halo_vary}).
Since the conversion probability $P(\gamma \rightarrow a) \propto |\textbf{B}|^2$, with ALPs being formed at every location, the ALP background power spectrum will vary as $D_{\ell} \propto {|\textbf{B}|^{4}}$ as shown in Fig.\ref{fig:neB_halo_vary}. A higher magnetic field leads to a higher probability of conversion at all locations in the galaxy cluster.

The polarized signal will be affected due to the scattering effects from regions of high-density electrons as the photon travels through the ICM of the clusters. Thus polarization information from high redshift clusters can be lost in the process. In our analysis, the primary contribution to the signal comes from low redshift (see Fig. \ref{fig:zmin_halo_vary} and Fig. \ref{fig:zmax_halo_vary} (in the appendix)). So, this effect will not impact the signal significantly. Another effect that can be important is the impact of the turbulent magnetic field and small-scale stochasticity on electron density. The presence of these effects will depolarise the signal and will produce an additional temperature fluctuation in the CMB due to ALPs \cite{Mukherjee_2018}. In the future, we plan to study this effect from cluster simulations and by jointly predicting the ALP signature in both temperature and polarization anisotropy. The ellipticity and geometry of the clusters can also play a role, but since the background signal will be an integrated effect over many clusters, the impact due to variation in the shape will be averaged out. However, the shape of the galaxy cluster will be important for the search of ALP signatures from resolved galaxy clusters \cite{mehta2024power}.

\section{Power spectrum estimation from sky maps} \label{sec:estimator}

The presence of one realization of the various signals requires us to define an estimator that takes into account the power spectrum for different components.
The ALP power spectrum can be estimated from the observed CMB power spectrum at the map level. 
So, we find the following estimator for any signal power spectrum using the maximum likelihood method \cite{Dodelson:2003ft} (explained in Appendix \ref{sec:estim_deri}): 
\begin{equation}
\tilde{C_{\ell}^{i}} = B_{\ell}^{-2} \left[ \frac{1}{2\ell + 1} \sum_{m = -{\ell}}^{{\ell}} |a_{\ell m}^{\mathrm{obs}}|^2 - N_{\ell} \right] - \sum_{j \neq i} C_{\ell}^{j}.
\label{eq:finestimator}
\end{equation}

The covariance of an estimator takes into account the limited information from an estimator at different angular scales due to the limited number of multipole modes (2$\ell + 1$) from a sky map. Also, it increases when a partial sky region is considered, as in our case of considering cluster regions. It is given as:
\begin{equation}
\mathrm{Cov}(\tilde{C_{\ell}}) = \langle \tilde{C_{\ell}}^2 \rangle - \langle C_{\ell} \rangle^2 = \frac{2}{(2\ell +1)f_{\mathrm{sky}}} [\sum_i C_{\ell}^i + B_{\ell}^{-2}N_{\ell}]^2 .
\label{eq:covar}
\end{equation}
The deviations in the CMB power spectrum due to the ALP distortion signal can be probed against the covariance for the null map spectrum to obtain bounds on the ALP background spectrum. The derivation of the estimator and its covariance are provided in Appendix \ref{sec:estim_deri}. 

%For a weighted combination of different frequency maps smoothed to common beam resolution, (see \cite{2014A&A...571A..21P}) applied in various cleaning techniques such as ILC, the estimator is given as:
%\begin{equation}
%\tilde{C_{\ell}^{i}} = B_{\ell}^{-2} \left[ \frac{1}{2\ell + 1} (\sum_{m = -l}^{l} \sum_{\nu}\sum_{\nu'} w_{\nu}w_{\nu'}a_{\ell m,\nu}^{obs\,*}   a_{\ell m,\nu'}^{obs} ) - \sum_{\nu} w_{\nu}^2 N_{\ell}^{\nu} \right] - \sum_{j \neq i} \sum_{\nu}\sum_{\nu'} w_{\nu}w_{\nu'}\langle a_{\ell m,\nu}^{j\,*} a_{\ell m,\nu'}^{j} \rangle.
%\label{eq:w-estimator}
%\end{equation}
%Here we have considered the independence of various components.

\section{Impact of foregrounds on the ALP signal} \label{sec:fgs impact}
\subsection{ALP background as compared to galactic foregrounds}\label{sec:fgs model}

The presence of foregrounds such as the thermal dust and synchrotron emission from the galactic plane impact the CMB polarization power spectrum substantially at low multipoles.  Their effect can be mitigated by masking the galactic plane and performing a partial sky observation. Masking is performed along the galactic plane to reduce the effect of foregrounds. This leads to a partial sky observation at high latitudes. The weak signal from the polarized SZ effect will be negligible of the order of tens of nano-Kelvins, hence we can neglect it \cite{Birkinshaw_1999,carlstrom2002cosmology}. 

\begin{figure}[h!]
     \centering
\includegraphics[height=8.5cm,width=14cm]{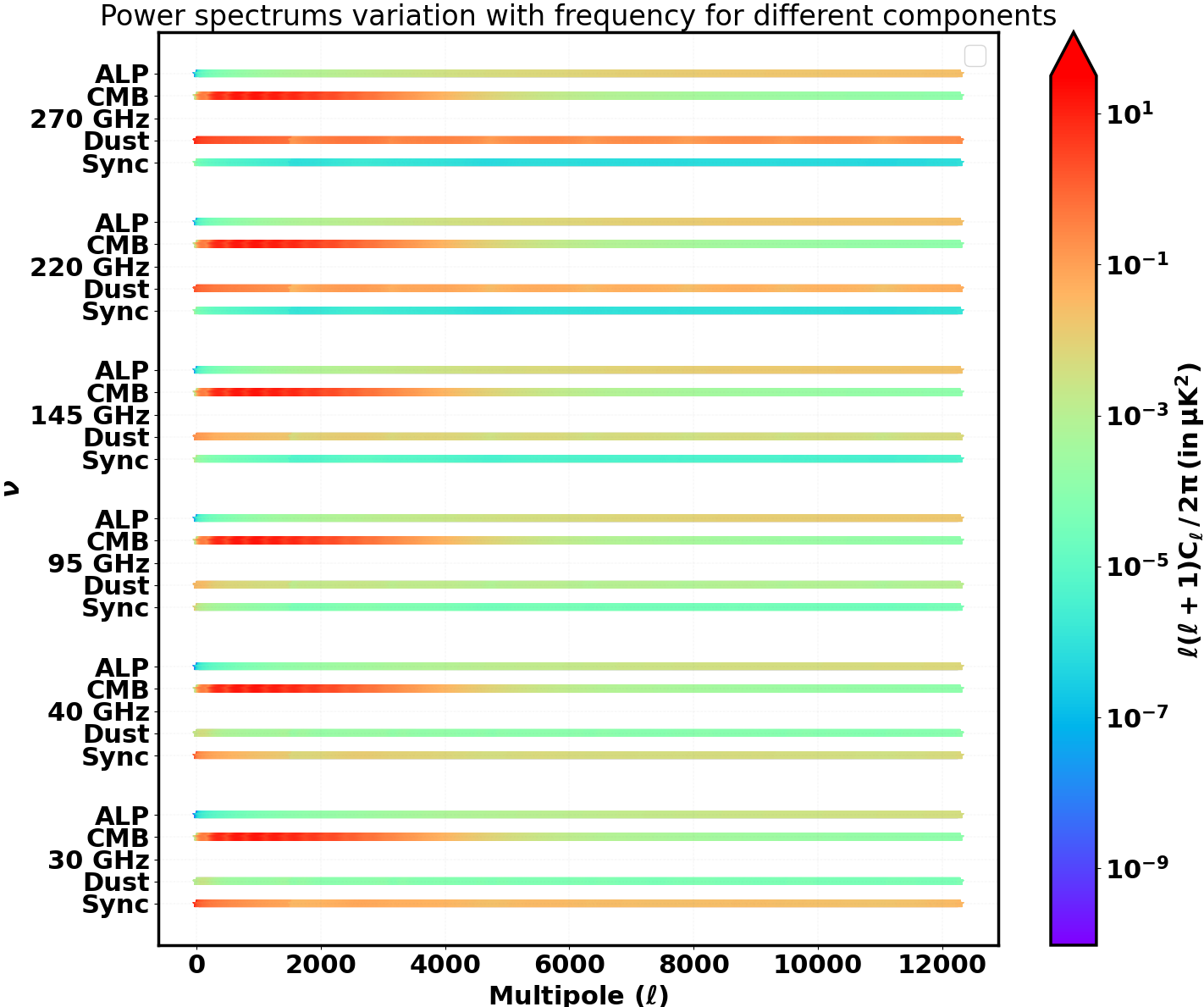}
    \caption{We show the spatial variation of the ALP background power spectrum ($g_{a\gamma} = 10^{-12} \, \mathrm{GeV^{-1}}$) at different multipoles ($\ell$) for various CMB-S4 frequency bands, as compared to CMB, galactic dust and galactic synchrotron emission. The CMB stays independent of frequency, while galactic synchrotron emission weakens with an increase in frequency. Dust and ALP background spectra increase with frequency, but they have a different multipole dependence, with dust power weakening at high multipoles, as opposed to ALP spectrum which increases at high multipoles. Also, dust follows the modified black-body model, while ALP spectrum simply scales as the squared of frequency.  }
       \label{fig:compare_plot}
\end{figure}

The thermal dust emission peaks at infrared wavelengths, hence it affects the power spectrum at high frequencies (200 GHz and above), while the synchrotron emission peaks at radio frequencies and its impact is maximum at low frequencies (below 70 GHz) (see Fig.\ref{fig:compare_plot}). At frequencies 90 - 170 GHz, the effect of the foregrounds is minimal.

Synchrotron emission weakens with an increase in frequency, while dust increases with frequencies in the microwave and radio regions of the EM spectrum (Fig.\ref{fig:compare_plot}). ALP diffused spectrum increases with frequencies, following a $D_{\ell} \propto \nu ^2 I_{\mathrm{cmb}}(\nu)^2$ dependence, as compared to dust which follows a modified black-body spectrum.
Both the galactic thermal dust and synchrotron emissions influence the power spectrum significantly at low multipoles, but at high multipoles, they weaken out, while the ALP diffused spectrum increases with multipoles following a $D_{\ell} \propto \ell^2$ dependence.  
These spatial and spectral variations of the ALP signal as compared to CMB and foregrounds can be used to detect the diffused ALP background spectrum.

\subsection{SNR for different CMB surveys}
The galaxy clusters are generated on the masked sky map in the unmasked regions. The sky fraction being observed depends on the detector, with sky fraction $f_{\mathrm{sky}} = 0.4$ for SO, while 0.5 for CMB-S4 and CMB-HD. The ALP distortion signal is simulated in these clusters. Beam smearing (denoted by $B_\ell$) occurs due to the resolution of the instrument and depends on the point source function. The combined map is then smeared with a Gaussian beam and instrument noise is added. The instrument noise for upcoming CMB surveys are taken assuming a Gaussian distribution. 

We check the detectability of the diffused spectrum using current and future detectors.
 The CAMB \cite{2011ascl.soft02026L} is used to generate the CMB power spectrum and map. 
 The SNR is calculated taking into account the contributions from the optimum multipole range, using both the polarized maps for various frequencies of observation. 
Thus, the squared signal-to-noise ratio (SNR, denoted by $\rho$) for the distortion signal power spectrum is found by summing over the contributions from Q and U polarized maps for the multipole range $\ell_{\mathrm{min}}$ to $\ell_{\mathrm{max}}$. We use the values of $\ell_{\mathrm{min}} = 1000$ and $\ell_{\mathrm{max}}$ corresponding to the beam resolution (multipole value at which $B_{\ell}^2 = 1 / e$) to obtain the SNR:
\begin{equation}
\rho_\nu^2 = \sum_{p = \mathrm{Q,U}} \sum_{{\ell}_{\mathrm{min}}}^{{\ell}_{\mathrm{max}}} \frac{(C_{{\ell},1h}^{p,\nu} + C_{{\ell},2h}^{p,\nu})^2}{\frac{2}{(2{\ell} + 1)f_{\mathrm{sky}}} (C_{{\ell},\mathrm{cmb}}^{p} + B_{{\ell},\nu}^{-2} N_{\ell}^{p,\nu})^2},
\label{eq:SNR2}
\end{equation}
where $C_{\ell}$ and $N_{\ell}$ are the true signal and noise power spectra respectively.

Here signal corresponds to the sum of one-halo and two-halo contributions, while the denominator is the covariance on the power spectrum of the observed sky. This is explained in Sec. \ref{sec:estimator}.
 The ${2\ell + 1}$ factor accounts for the number of modes for every multipole $\ell$. 

\begin{figure}[h!] 
     \centering
\includegraphics[height=5.5cm,width=11cm]{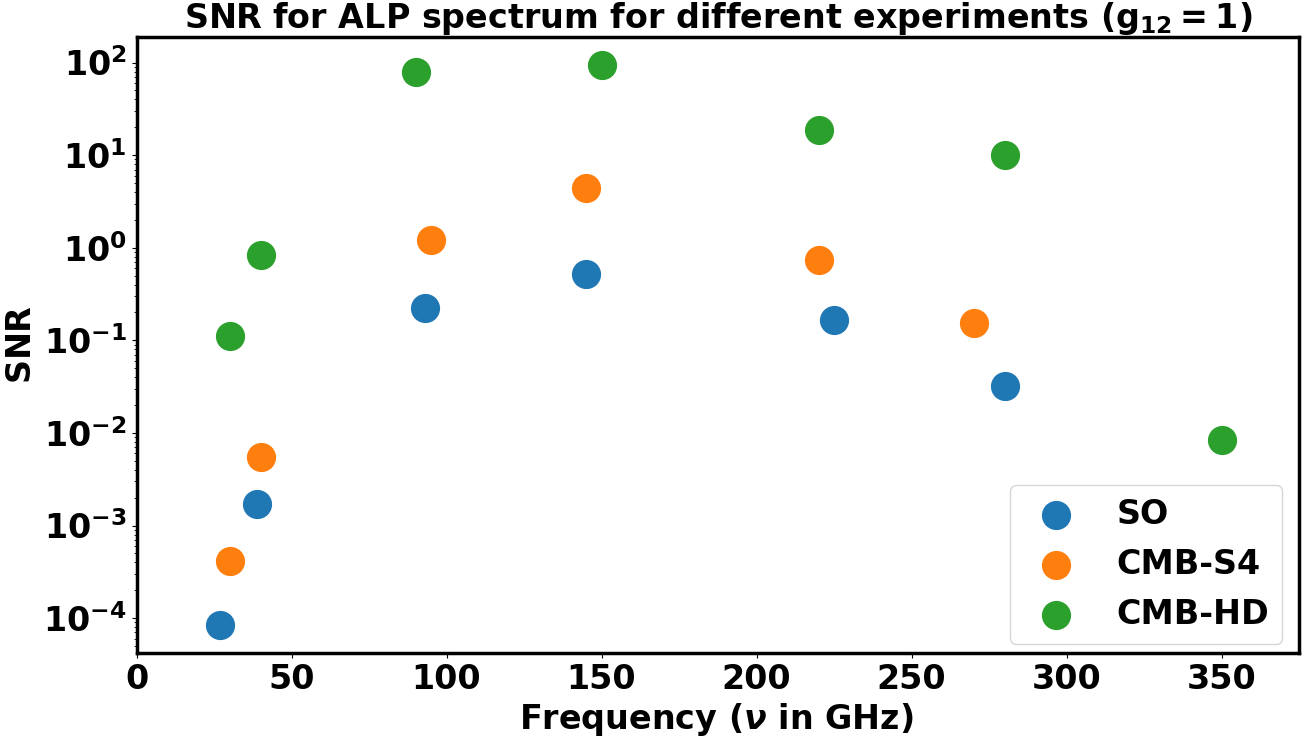}
   \caption{This figure shows the SNR obtained at various frequency bands for different CMB detectors, corresponding to a coupling constant of $g_{a\gamma} = 10^{-12} \, \mathrm{GeV^{-1}}$. The maxima for each detector is obtained for frequencies around 140 - 150 GHz, due to low beam size and instrumental noise. The presence of foregrounds reduces the SNR, but their impact is minute at frequencies 95 and 145 GHz, due to the weakening of the foregrounds in this frequency range.}
      \label{fig:snrEXP} 
\end{figure}
  
Upcoming experiments such as the CMB-S4 and CMB-HD could be particularly useful in estimating this power spectrum with a lower beam and instrument noise. With regards to the current detectors, a coupling strength of $10^{-12} \, \mathrm{GeV ^{-1}}$ with SO \cite{Ade_2019} configuration generates a maximum SNR of $\sim$ 0.52 (Fig. \ref{fig:snrEXP}). The SNR achieves the highest value of 4.36 at the 145 GHz band in the presence of foregrounds for CMB-S4 \cite{abazajian2016cmbs4}, due to low noise and high resolution. An SNR of $\sim$ 93.87 will be achievable with CMB-HD \cite{sehgal2019cmbhd} in the 150 GHz band. The presence of galactic foregrounds reduces the SNR, especially for low (impacted by synchrotron) and high frequency channels (impacted by dust). The bands from 90 to 170 GHz face the minimum impact as both dust and synchrotron weaken in this range of frequencies.     
 The SNR varies as $g_{a\gamma}^4$ with the coupling constant.

\section{ILC to extract axion signal from multi-frequency maps}\label{sec:ilc}
The ALP signal, contaminated with foregrounds and noise, accompanied by beam smearing, is difficult to extract. 
The spectral shape of the ALP background signal can be used to clean contaminants like CMB and foregrounds.
  The standard Internal Linear Combination (ILC) may be used to obtain a clean map as it minimizes the variance of the map and extracts a given spectrum. 
   The higher the number of frequencies, accompanied by a lower beam size, the better the results obtained after ILC. 

The ILC \cite{Eriksen_2004,ilc2008internal} is applied to extract the ALP-distortion signal. The method combines maps from different frequencies and assigns weights to them depending on the spectrum to be obtained (Fig. \ref{fig:compare_plot}). The weighted sum is then the ILC map with the required spectrum. The weights are obtained using the covariance matrix which combines data at different frequencies. The weight matrix for standard ILC is given as:
\begin{equation}
    w_{\mathrm{ilc} } = f_{a \gamma}^T C_s^{-1} (f_{a \gamma}^T C_s^{-1} f_{a \gamma})^{-1}    
    .
    \label{eq:ilc w}
\end{equation}
Here $f_{a \gamma}$ is the ALP distortion spectrum dependence on the frequency with $f_{a\gamma} \propto \nu I_{\mathrm{cmb}}(\nu)$. The ILC cleaned map is obtained as a weighted linear combination of the frequency maps ($S^{\nu}$):
\begin{equation}
    S_{\mathrm{ilc}} = \sum_{\nu} w_{\mathrm{ilc}}^{\nu} S_{\nu} .
    \label{eq:ilcmap}
\end{equation}

The distortion signal is simulated in galaxy clusters at different frequencies with a constant coupling constant ($g_{a\gamma}= \mathrm{1 \times 10^{-12} \, GeV^{-1}}$ here) for all clusters. The signal is then contaminated with CMB and galactic foregrounds. 
We have used PySM \cite{Thorne_2017} to generate foreground maps.
We use the synchrotron model "s-3" model which includes a  curved index that flattens or steepens with frequency and dust model "d-3", which takes into account the spatial variation of spectral index on degree scales. The index is drawn from a Gaussian distribution.
 The various frequency maps are smoothed to a common beam resolution, given by the highest beam size among various frequency bands. The bands above 200 GHz can efficiently clean thermal dust contribution in the cleaned map. The CMB is frequency-independent, while the synchrotron emission is very weak above a frequency of 70 GHz. This lets us use just four bands with a higher beam resolution at frequencies greater than 90 GHz. Masking is done in galactic regions. The ALP signal dominates at high multipoles, owing to the one-halo term, while the foregrounds and CMB dampen out. It is the multipole range around $\ell = 3000$ that can provide the best constraints on the ALP background spectrum, because at very high multipoles the noise takes over. Using ILC, we get the following SNRs for the various detectors for the ALP diffused spectrum with $z_{\mathrm{min}} = 1$:  0.24 with SO; 1.20 with CMB-S4; 79.27 with CMB-HD. This SNR is the best we can achieve using ILC when the foregrounds are not well known. If they are well modelled, we can use template matching to achieve a higher SNR at the matched frequency bands corresponding to 140-150 GHz.

The ALP signal will be correlated with other weak signals like the CIB emission and polarized SZ effect. This correlation may affect the efficiency of the ILC cleaning method. To deal with these, we can also make use of methods like the constrained ILC (both for temperature and polarization fluctuations) \cite{Remazeilles_2010,Remazeilles_2021,zhang2024constrained}, although at a noise tradeoff, which can remove contaminations with a given spectrum. It can make use of the spectral information of the contaminant to completely or partially eliminate it. We assume in our analysis that the effect of these contaminants is low and the spectral and spatial shape of the ALP spectrum can distinguish the signal from these contaminants.  

\section{Constraining the ALP coupling constant using ILC}\label{sec:constraints}

We perform the Bayesian estimation of the photon-ALP coupling constant from diffused ALP spectrum. The background spectrum scales as $g_{a\gamma}^4$ (Fig.\ref{fig:ga_halo_vary}). Since the electron density and magnetic field information for the unresolved clusters is lacking, and from Sec.\ref{sec:ne_B_halo}, we know that the background spectrum scales as $n_e^{-2}$ and $|\textbf{B}|^4$
(Fig.\ref{fig:neB_halo_vary}), we thus obtain bounds on the quantity which we will call the modified coupling constant: $g_{a\gamma}^2|\textbf{B}|^2 / n_e$. The fiducial values we have used for the electron density and magnetic field strength parameters are: $n_0 = 10^{-3} \, \mathrm{cm^{-3}}$ and $B_0 = 0.1 \, \mathrm{\mu G}$. We scale the modified coupling constant with respect to this choice of values to obtain bounds on the scaled quantity given as:
\begin{equation}\label{eq:A eq}
    A = \left(\frac{g_{a\gamma}}{10^{-12} \, \mathrm{GeV^{-1}}} \right)^2 \left( \frac{|\textbf{B}|}{0.1 \, \mathrm{\mu G}} \right)^2 \left( \frac{n_e}{10^{-3} \, \mathrm{cm^{-3}}} \right) ^{-1}.
\end{equation}

Since the ALP background spectrum depends significantly on the minimum redshift $z_{\mathrm{min}}$ and minimum cluster mass $M_{\mathrm{min}}$, we obtain constraints on the modified constant for $z_{\mathrm{min}} = 0.5$ and $z_{\mathrm{min}} = 1$. For a given $z_{\mathrm{min}}$, we show the bounds for different choices of $M_{\mathrm{min}}$. We vary $A$ with the upper bound set by the values: $g_{a \gamma} = 10^{-11} \, \mathrm{GeV^{-1}}$, $n_0 = 0.5 \times 10^{-3} \, \mathrm{cm^{-3}}$ and $B_0 = 0.5 \, \mathrm{\mu G}$.
We combine the posteriors from different mass ranges to obtain the bounds for a given choice of $z_{\mathrm{min}}$.

We compare the bounds on modified coupling constant obtained using ILC for different CMB surveys: SO \cite{Ade_2019}, CMB-S4 \cite{abazajian2016cmbs4} and CMB-HD \cite{sehgal2019cmbhd}. Along with these experiments, dedicated CMB spectral distortion experiments such as the Primordial Inflation Explorer (PIXIE) \cite{kogut2011primordial,kogut2016primordial,kogut2024primordial} with high sensitivities and a high spectral resolution can also be used for measuring the coupling constant of ALPs using diffused background signal if the distortion polarization information is well resolvable. This will require a high beam resolution, otherwise the signal will be depolarized due to convolution with the beam. Future CMB experiments with an appropriate design with both spectral and spatial resolution can bring new discoveries in this area.

We simulate the fiducial maps at various frequencies without ALP signal ($g_{a\gamma} = 0$). We obtain the weights for these maps using Eq.\ref{eq:ilc w} and linearly combine them with their respective weights to obtain the ILC map. This method extracts the ALP signal, while minimizing the variance of the ILC map. We obtain the power spectrum of this map which gives us the term $\sum_{m = -\ell}^{\ell}|a_{lm}^{\mathrm{obs}}|^2$ in Eq.\ref{eq:finestimator} at different multipoles. Since noise for different frequency maps is not correlated, we obtain the term $N_{\ell}$ in Eq.\ref{eq:finestimator} by combining weighted noise maps at various frequencies. Also, we obtain the mean power spectrum of different realizations of the combined fiducial ILC map (without ALP signal) by combining the fiducial maps at different frequencies (based on their ILC weights)  to obtain the term $ C_{\ell}^{\mathrm{cmb}} + C_{\ell}^{\mathrm{fg}}$. These give us an estimation of the ALP background spectrum  $C_{\ell}^{\mathrm{ax}}$. The beam $B_{\ell}$ corresponds to the maximum beam size among the various frequency bands, as all maps are smoothed with the highest beam size before ILC weights are obtained.  

  \begin{figure}[h!]
     \centering
\includegraphics[height=6cm,width=11cm]{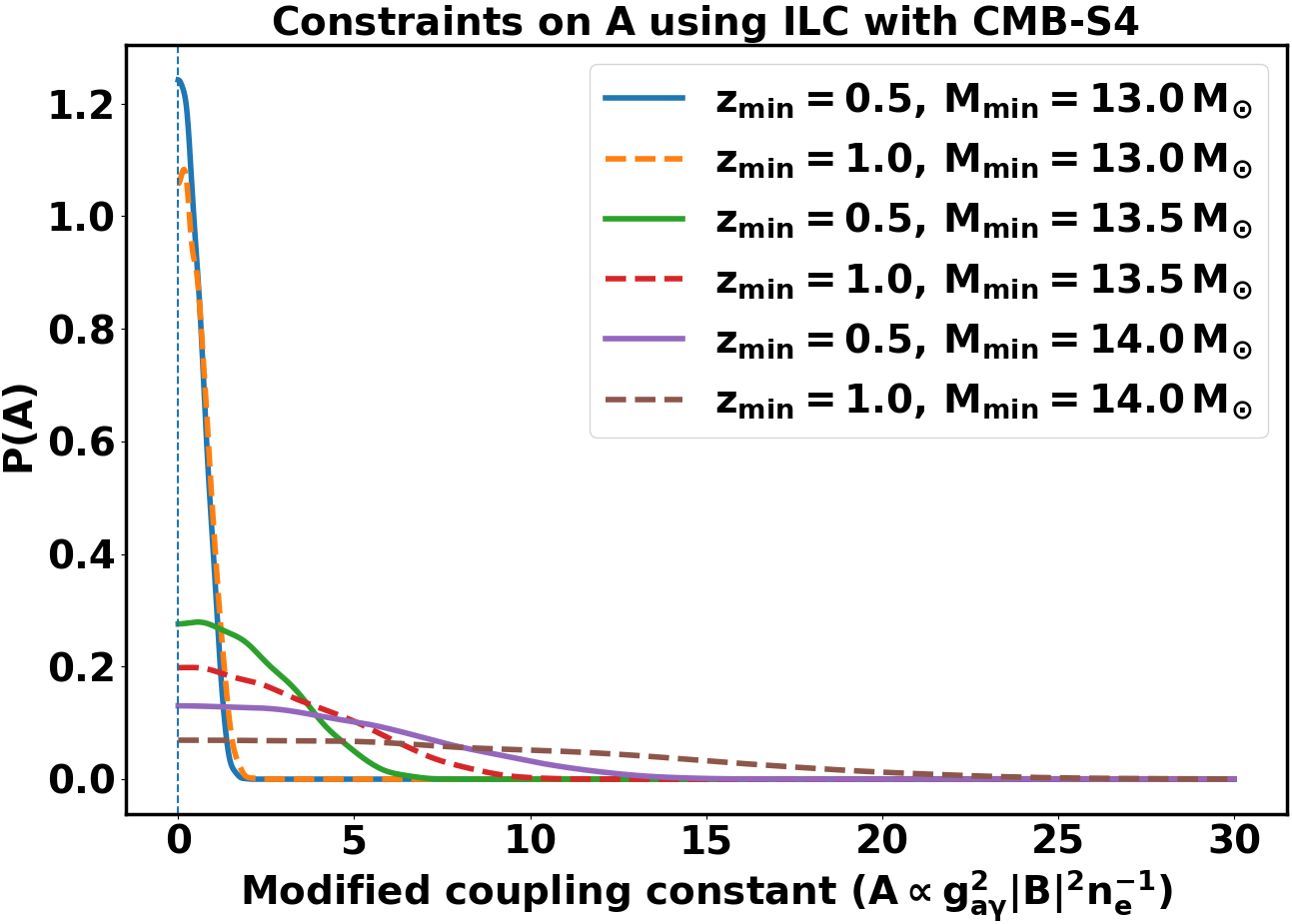}
    \caption{Constraints from CMB-S4 using ILC for $z_{\mathrm{min}} = 0.5$ and $z_{\mathrm{min}} = 1$}
       \label{fig:s4_ilc}
\end{figure}
We use 50\% partial sky to obtain bounds using CMB-S4 configuration, with frequency bands 95, 145, 220 and 270 GHz with the beam 2.2 arcmin corresponding to 95 GHz band. The bounds are stronger in case of a lower $M_{\mathrm{min}}$ as can be seen in Fig.\ref{fig:s4_ilc}. In principle the improvement when using a lower $M_{\mathrm{min}}$ will depend on the relation between mass and electron density of low mass galaxy clusters, which will determine their contribution to the ALP background. 

The constraints are also tighter in the case of 
$z_{\mathrm{min}} = 0.5$  as compared to $z_{\mathrm{min}} = 1$ (Fig.\ref{fig:s4_ilc}), as expected with low redshift clusters contributing to the background. Combining the posteriors for different mass ranges, we obtain the bounds on modified coupling constant using ILC with CMB-S4 configuration up to 95\% confidence interval at $A < 1.116$ for $z_{\mathrm{min}} = 0.5$, while $A < 1.254$ for $z_{\mathrm{min}} = 1$.
  \begin{figure}[h!]
     \centering
\includegraphics[height=6cm,width=11cm]{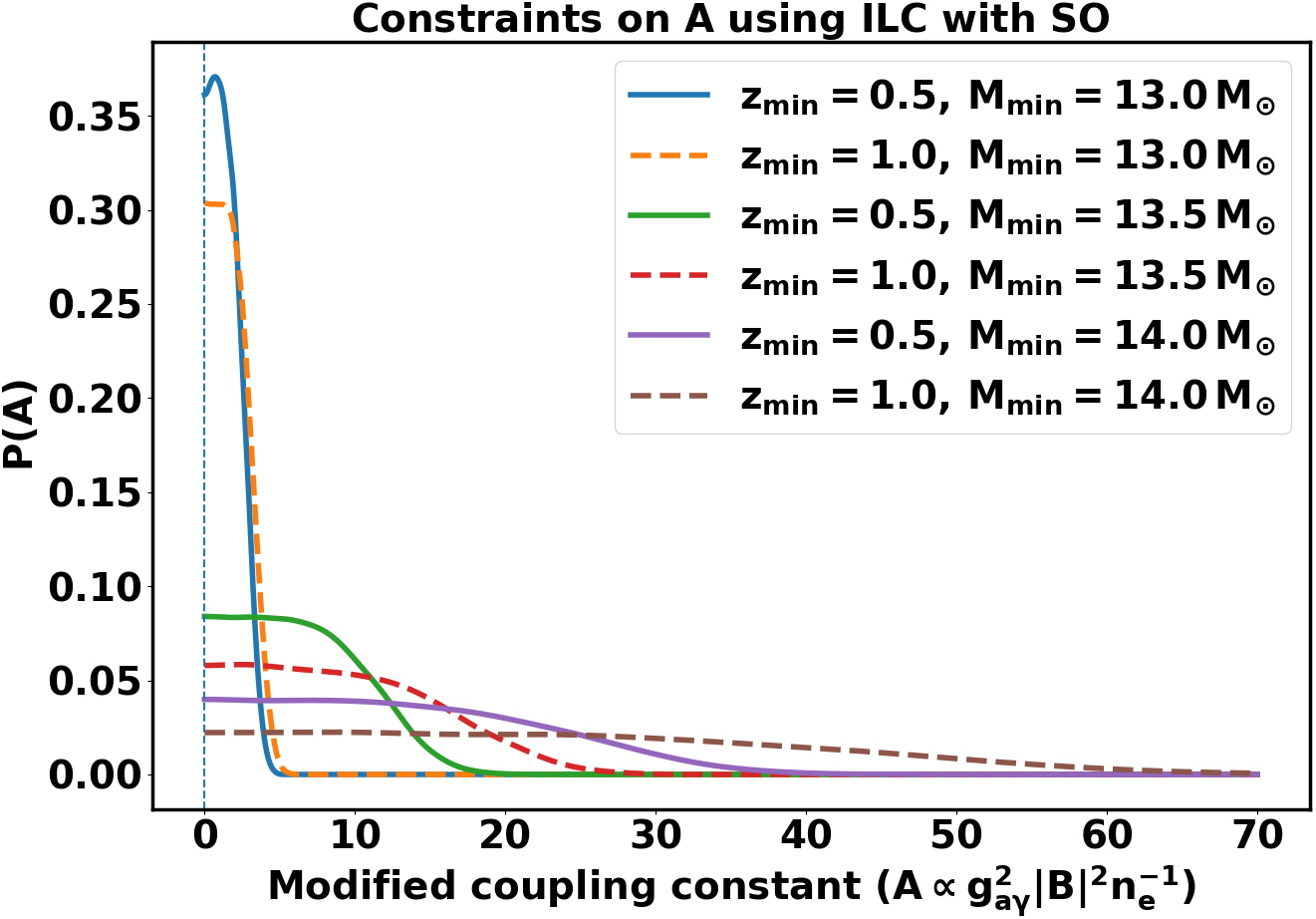}
    \caption{Constraints from SO using ILC for $z_{\mathrm{min}} = 0.5$ and $z_{\mathrm{min}} = 1$}
       \label{fig:so_ilc}
\end{figure}

For SO bounds, we use 40\% sky fraction. The frequency bands 93, 145, 225, 280 GHz are used with the common beam 2.2 arcmin corresponding to the 93 GHz band. The constraints are shown in Fig.\ref{fig:so_ilc}. We obtain the following bounds using ILC, up to 95\% confidence interval: $A < 3.178$ for $z_{\mathrm{min}} = 0.5$;  $A < 3.724$ for $z_{\mathrm{min}} = 1$.

  \begin{figure}[h!]
     \centering
\includegraphics[height=6cm,width=11cm]{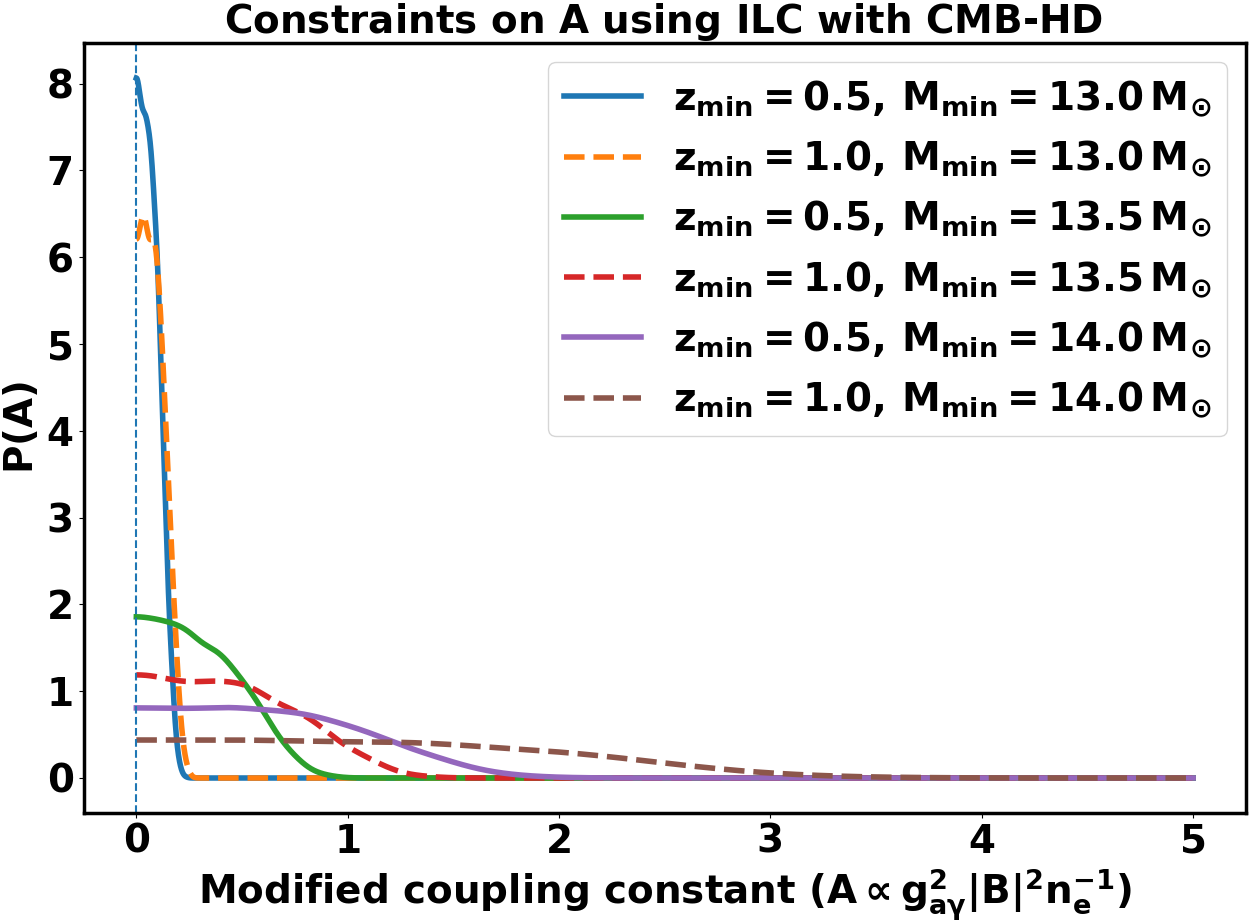}
    \caption{Constraints from CMB-HD using ILC for $z_{\mathrm{min}} = 0.5$ and $z_{\mathrm{min}} = 1$.}
       \label{fig:hd_ilc}
\end{figure}
CMB-HD will have a partial sky fraction of 50\%. We use the 90, 150, 220 and 280 GHz frequency bands to obtain bounds on the modified coupling constant with the common beam resolution of 0.42 arcmin, corresponding to the band 90 GHz. The results are shown in  Fig.\ref{fig:hd_ilc}. Using ILC, the bounds on modified coupling constant upto 95\% confidence interval (C.I.) are: $A < 0.115$ for $z_{\mathrm{min}} = 0.5$;  $A < 0.131$ for $z_{\mathrm{min}} = 1$. CMB-HD will provide bounds significantly better than CMB-S4 and SO on the modified coupling constant due to a higher beam resolution and improved sensitivity.
The constraints obtained using the template matching of foregrounds are discussed in Sec.\ref{sec:tempmatch}.

\begin{figure}[h!]
     \centering
\includegraphics[height=7cm,width=11cm]{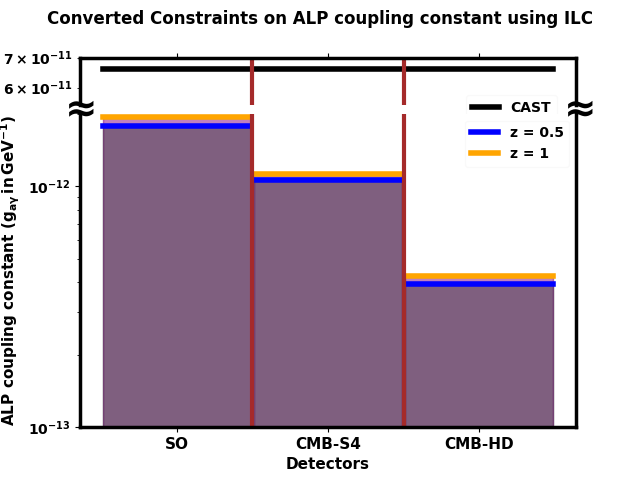}
    \caption{Converted bounds on ALP coupling constant for various surveys with $z_{\mathrm{min}} = 0.5$ and $z_{\mathrm{min}} = 1$}
       \label{fig:ilcconv}
\end{figure}

The constraints on photon-ALP coupling constant $g_{a\gamma}$ can be gauged using the constraints on the modified coupling constant $A$. We convert the constraints on $A$ to the constraints on $g_{a\gamma}$ using the fiducial values of the magnetic field and electron densities used to obtain $A$, i.e. $n_0 = 10^{-3} \, \mathrm{cm^{-3}}$ and $B_0 = 0.1 \, \mathrm{\mu G}$. These are plotted in Fig.\ref{fig:ilcconv} with the horizontal lines representing the 95\% C.I. bounds for various detectors, with the allowed coupling strengths shaded.  We expect an upper bound on ALP coupling strength of $g_{a\gamma} < 1.929 \times 10^{-12} \, \mathrm{GeV^{-1}}$ using SO and
of about $g_{a\gamma} < 1.119 \times 10^{-12} \, \mathrm{GeV^{-1}}$ using CMB-S4 configurations for $z_{\mathrm{min}} = 1$. ILC using CMB-HD can strongly constrain the coupling to $g_{a\gamma} < 4.254 \times 10^{-13} \, \mathrm{GeV^{-1}}$.

The constraints will also depend on the ALP masses that are being considered. For higher mass ALPs ($10^{-13} - 10^{-11}$ eV), the constraints will be weaker as compared to low mass ALPs ($10^{-15} - 10^{-13}$ eV). Since we have considered ALPs of masses in the range $10^{-15} - 10^{-11}$ eV, we have obtained the strongest constraints for the configuration using the ALP diffused spectrum.

\section{Conclusion}\label{sec: conclusion}
The galaxy clusters are the largest visible gravitationally bound structures. If ALPs exist in the universe and weakly interact with photons, the CMB will bear the polarized distortion spectrum from these clusters. The ALP signal from resolved clusters can be independently studied\cite{mehta2024power,Mehta:2024:new3,Mukherjee_2020}, while the unresolved ones would create a polarized ALP background signal.

This background signal will depend on the cluster mass distribution at different redshifts, which are biased tracers of the dark matter halos. This is taken into account using the distribution of these halos and the correlation between different halos. The background will be an integrated effect of the signals from individual clusters of various masses at different redshifts. This diffused spectrum can be modelled using the distribution of halos of various masses at different redshifts. The two-halo or clustering component is low at high multipoles, but may dominate over the one-halo term or the Poissonian component for low ones (20 to 200), where it is itself  dominated by the CMB and is difficult to probe. The one-halo or Poissonian component of the ALP power spectrum dominates at high multipoles and can be probed using the upcoming high resolution experiments (CMB-S4, CMB-HD) with low noise. The ALP background  spectrum will depend on the astrophysical aspects like the electron density and magnetic field profiles in clusters (see Figs. \ref{fig:neB_halo_vary}). Also, it will increase with  ALP coupling constant ($\propto g_{a\gamma}^4$) (see Fig. \ref{fig:ga_halo_vary}) and the frequency of observation ($\propto \nu ^2 I_{\mathrm{cmb}}(\nu)^2$)(see Fig. \ref{fig:compare_plot}). 

The background spectrum will increase as we lower $z_{\mathrm{min}}$, as low redshift clusters contribute significantly to the background spectrum (Fig. \ref{fig:zmin_halo_vary}). With the upcoming experiments, clusters up to redshift $z = 1$ will be well resolvable, thus $z_{\mathrm{min}} = 1$ is a conservative choice for the estimation of the ALP background spectrum. The power spectrum is almost independent of the choice of maximum redshift $z_{\mathrm{max}}$ after a certain redshift ($\sim 3.5$) as the clusters at very high redshifts contribute negligibly to the ALP background spectrum (Appendix \ref{sec:z_vary}). The cluster mass range that will contribute to the background will also determine the strength of the diffused power spectrum (Fig.\ref{fig:m_halo_vary}). 
The ALP masses that may exist will also affect the background spectrum, with a weaker power spectrum for high mass ALPs.

For a coupling constant of $10^{-12} \, \mathrm{GeV^{-1}}$ and $z_{\mathrm{min}} = 1$, with randomly generated cluster profiles for cluster masses $\mathrm{10^{13} - 7 \times 
 10^{15} \, M_{\odot}}$, the SNR is 4.36 in the 145 GHz band of CMB-S4, while it is around 93.87 in the 150 GHz band of CMB-HD (Fig. 
 \ref{fig:snrEXP}). Also, such a diffused signal would lead to RMS fluctuations of the order of $\mathrm{7.5 \times 10^{-2} \, \mu K}$ for an ALP coupling constant of $g_{a\gamma} = 10^{-12} \, \mathrm{GeV^{-1}}$ at 145 GHz. The frequency channels 90 - 160 GHz provide the best SNR for the ALP signal, owing to a decrease in foregrounds contamination and improved beam resolution. 

Techniques such as ILC (see Sec.\ref{sec:ilc}) can be used to mitigate the effect of foregrounds and CMB by using the spectral variation of the ALP signal (Fig.\ref{fig:compare_plot}). Using ILC for different detectors, we obtain the following bounds on the modified coupling constant $A$ (see Sec.\ref{sec:constraints}):
$$\mathrm{SO}: A < 3.724; $$ $$\mathrm{CMB-S4}: A < 1.254;$$
$$ \mathrm{CMB-HD}: A < 0.181,$$     
    
for the case of $z_{\mathrm{min}} = 1$. The converted constraints on the coupling constant $g_{a\gamma}$ can be gauged using the fiducial electron density and magnetic field strength values. The conversion provides the bounds of $ g_{a\gamma} < 1.783 \times 10^{-12} \, \mathrm{GeV^{-1}}$ with SO and
 $g_{a\gamma} < 1.056 \times 10^{-12} \, \mathrm{GeV^{-1}}$ with CMB-S4 configuration for $z_{\mathrm{min}} = 0.5$. CMB-HD can provide much tighter bounds of $g_{a\gamma} < 3.912 \times 10^{-13} \, \mathrm{GeV^{-1}}$. For $z_{\mathrm{min}} = 1$, the bounds go as: 
 $$\mathrm{SO} : g_{a\gamma} < 1.929 \times 10^{-12} \, \mathrm{GeV^{-1}};$$  $$\mathrm{CMB-S4} : g_{a\gamma} < 1.119 \times 10^{-12} \, \mathrm{GeV^{-1}};$$ 
 $$\mathrm{CMB-HD }: g_{a\gamma} < 4.254 \times 10^{-13} \, \mathrm{GeV^{-1}}.$$ 

 The template matching of foregrounds can tighten the constraints on coupling constant as shown in Appendix \ref{sec:tempmatch}. The shape of the foregrounds power spectrum and their violation of statistical isotropy can be used to reduce further contamination of foregrounds. This can help in achieving a higher SNR and better constraints. The galaxy clusters, with their intracluster medium (ICM) and galaxies make sites for numerous processes such as the SZ effects, lensing, CIB, synchrotron, etc. The ALP power spectrum can also be cross-correlated with diffused spectra from these phenomena and the large-scale structure  to obtain better constraints on the ALP coupling constant and masses \cite{mondino2024axion}. 

In this analysis, we have neglected the effect of evolution of galaxy clusters with redshift. Our results depend on the consideration of random electron density and magnetic field profiles for the clusters. 
Due to the low redshift clusters being way more than the high redshift clusters, most of the contribution to the calculated background comes from low redshift clusters which have higher halo mass function values and for which the polarization signal remains intact. Thus, a study of the evolution of the cluster profiles with masses at different redshifts can provide bounds on this diffused spectrum.
Not only this, being able to connect the masses of galaxy clusters and their electron density profiles would further help in constraining the photon-ALP coupling constant. This can be done (as shown in the case of SZ effect by \cite{Komatsu2000CMBAF}) using hydrodynamical simulations like Romulus \cite{Tremmel_2017}, SIMBA \cite{Dav__2019}, etc. 

Probing the high multipoles with improved detectors will help in obtaining bounds on this diffused signal in the future. The background ALP sky can thus, in addition to probing the signal from resolved clusters, act as an independent way of obtaining constraints on the ALP coupling constant, by studying its spectral and spatial variation over a wide range of frequencies and multipoles.

\appendix
\section{The ALP distortion power spectrum for a single cluster $\mathrm{| \alpha _{\ell}|^2}$}. \label{sec:alpha_l}
The photon-ALP conversion is confined to galaxy clusters which occupy small angular scales on the sky. 
For a particular ALP mass range, the signal is formed in a spherical shell around the cluster center for a spherically symmetric galaxy cluster. These shells are projected as a disk in 2d. This signal disk increases the ALP power spectrum at low angular scales.
It is the higher multipoles that contain ALP signatures in the CMB spectrum as multipoles vary inversely with the angular scales $\mathrm{\Delta \theta \propto 180^o / \ell}$. 

The $\alpha_{\ell}$'s are the coefficients obtained from the  spherical harmonics expansion of the polarization fluctuations caused due to the photon-ALP conversion. The $|\alpha_{\ell}|^2$ is the power spectrum of these polarization fluctuations due to the ALP distortion signal. The estimation of this power spectrum due to temperature or polarization fluctuations in a map is explained in Appendix \ref{sec:estim_deri}.

The astrophysics of galaxy clusters affects their electron density and magnetic field profiles, and hence the ALP distortion spectrum $|\alpha_{\ell}|^2$. For clusters with stochastic and turbulent electron density and magnetic field profiles, the polarization information of the ALP distortion will be lost as depolarization will be caused due to multiple conversions. For such cases, hydrodynamical simulations or observations can be used that fit the data well. We consider the case where this turbulence is negligible and model the electron density and magnetic profiles using smooth profiles, which fit the profiles for resolved clusters well.

The electron densities can be obtained via X-ray emission and inverse-Compton (Sunyaev-Zeldovich) effect in galaxy clusters. The electron density profile used is a modified beta model that varies radially and takes into account the slope at large radii, a cusp core that follows a power law, and the higher electron density in the inner regions \cite{Vikhlinin_2006,mcdonald2013growth}:
\begin{equation}
  n_e^2 = Z\left[ n_0^2 \frac{(r/r_{c1})^{-\alpha}}{(1 + r^2/r_{c1}^2)^{3\beta - \alpha /2}}\frac{1}{(1 + r^{\gamma}/r_s^{\gamma})^{\epsilon / \gamma}} + \frac{n_{02}^2}{(1 + r^2/r_{c2}^2)^{3\beta_2}}\right].
\label{eq:elec dens}\end{equation}
The photon-ALP resonant conversion depends on the transverse magnetic field along the line of sight. The transverse magnetic field profile can be modelled using synchrotron emission by studying the galaxy cluster at radio frequencies.
We consider a magnetic field profile (that models well for clusters at low redshifts) whose strength simply scales with distance from the cluster center \cite{Carilli_2004,bonafede2010galaxy,bohringer2016cosmic}:
 \begin{equation}
   B(r) = B_0 r^{-s}.
\label{eq:mag prof}
\end{equation}
The contribution to the Q and U polarization stokes modes depends on the magnetic field direction at the conversion location. 
The magnetic field direction has been assumed to be uniformly randomly oriented. 
The magnetic field coherence scale is assumed to be greater than the beam of the instrument, otherwise the signal will be depolarized by the turbulent fields. The power spectrum $|\alpha_{\ell}|^2$ is that of the combined power in the two maps for which the polarized intensity  $I_{\mathrm{pol}}$ is given as:
\begin{equation}
I_{\mathrm{pol}} = \sqrt{I_{\mathrm{Q}}^2 + I_{\mathrm{U}}^2}. 
\end{equation}

\begin{figure}[h!]
     \centering
     \begin{subfigure}[h]{0.45\textwidth}
         \centering    
\includegraphics[height=5cm,width=7cm]
         {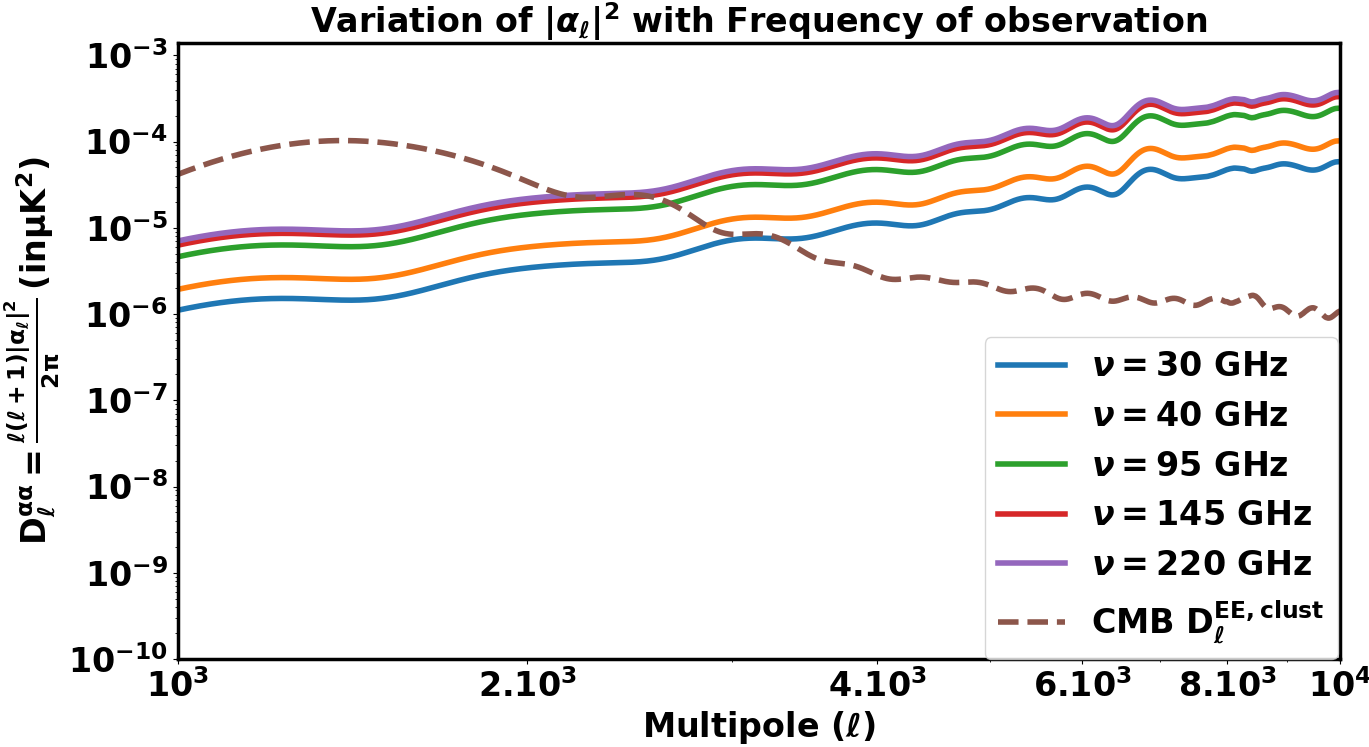}
         \caption{Variation in $\mathrm{|\alpha_{\ell}|^2}$ with frequency of observation ($\propto \nu^2$)}
         \label{fig:f_alp_vary}
     \end{subfigure} 
     \hspace{0.01cm} 
     \begin{subfigure}[h]{0.45\textwidth}
         \centering
    \includegraphics[height=5cm,width=7cm]
         {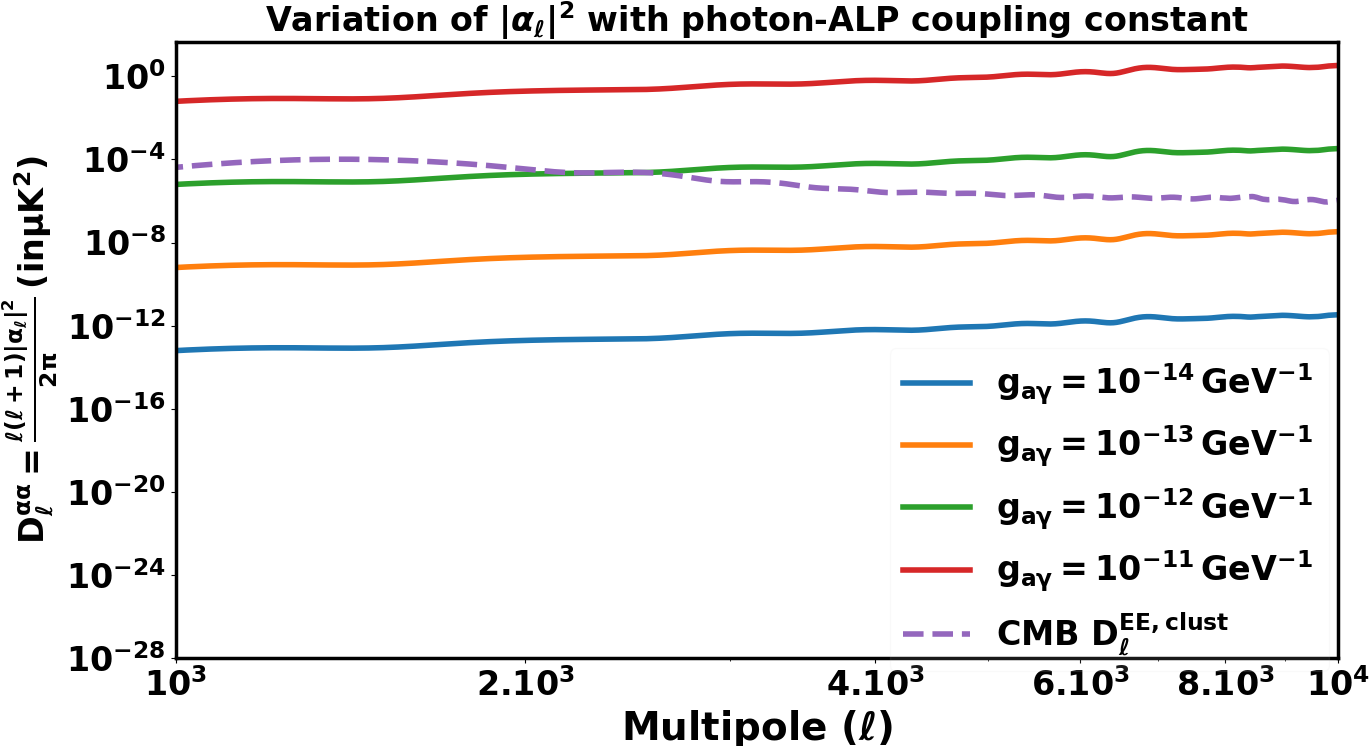}
         \caption{Variation in $\mathrm{|\alpha_{\ell}|^2}$ with coupling constant ($\propto \mathrm{g_{a\gamma}}^4$)}
         \label{fig:g_alp_vary}
     \end{subfigure} 
\caption{Variation in $\mathrm{|\alpha_{\ell}|^2}$ with frequency (in the microwave and radio spectrum) and ALP coupling constant.  The CMB polarization power spectrum at small scales (around the cluster region) is
also plotted.}
\label{fig: alpha_f_g}
\end{figure}

The power spectrum $|\alpha_{\ell}|^2$ will also vary for the mass range of ALPs being considered. 
ALPs of masses in a particular sub-range of the mass range $10^{-15} - 10^{-11}$ eV may be assumed to be forming in the galaxy clusters if the resonant condition is satisfied ($m_a = m_{\gamma} $). For our analysis, we assume that if ALPs exist, all ALPs of masses in the range $10^{-15} - 10^{-11}$ eV will be produced during conversion. Also, we assign a uniform coupling constant of $\mathrm{10^{-12} \, GeV^{-1}}$ for all ALP masses in this range. 
\begin{figure}[h!]
     \centering
     \begin{subfigure}[h]{0.45\textwidth}
         \centering    
\includegraphics[height=5cm,width=7cm]
         {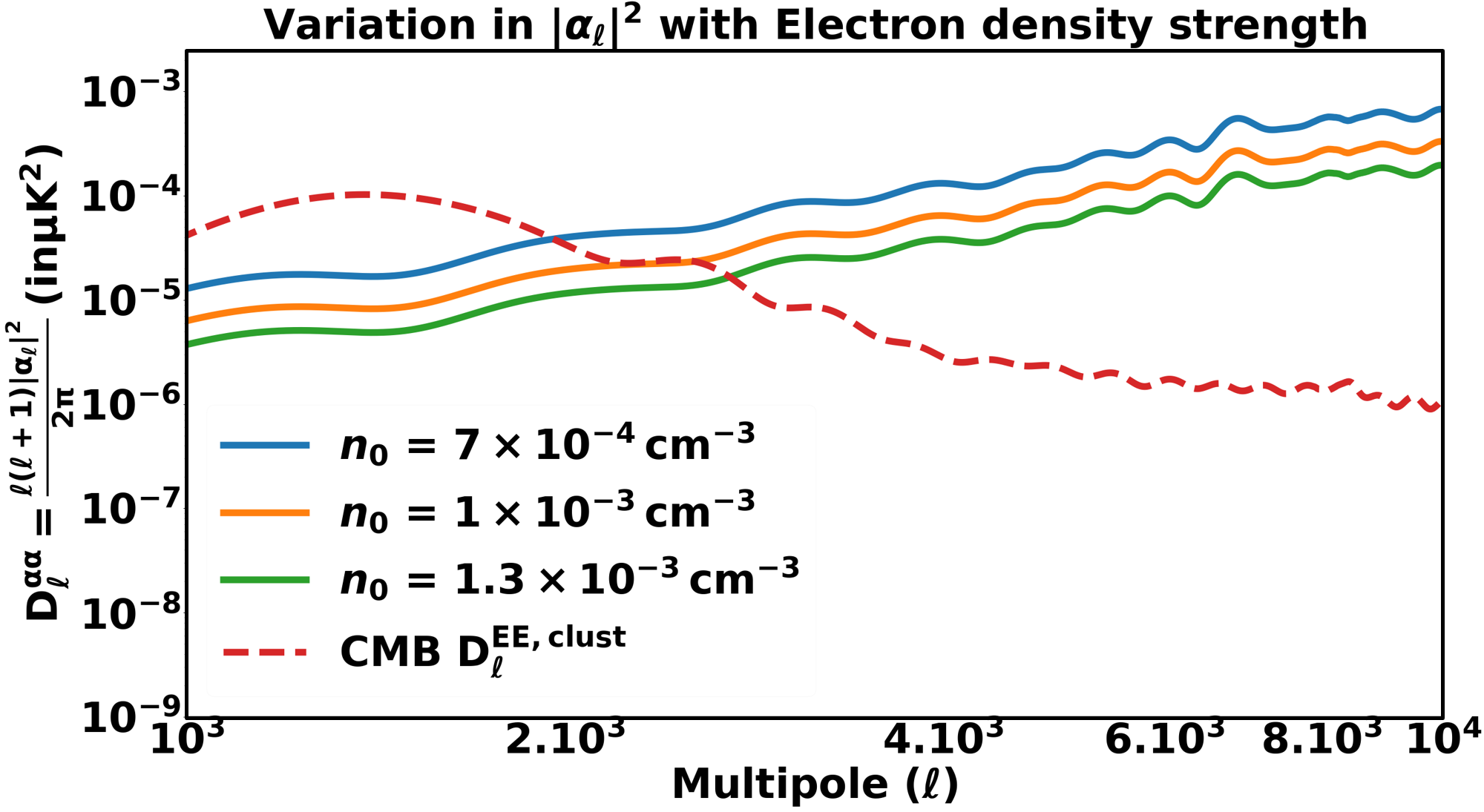}
         \caption{Variation in $\mathrm{|\alpha_{\ell}|^2}$ with Electron density strength}
         \label{fig:n0_alp_vary}
     \end{subfigure} 
     \hspace{0.01cm} 
     \begin{subfigure}[h]{0.45\textwidth}
         \centering
    \includegraphics[height=5cm,width=7cm]
         {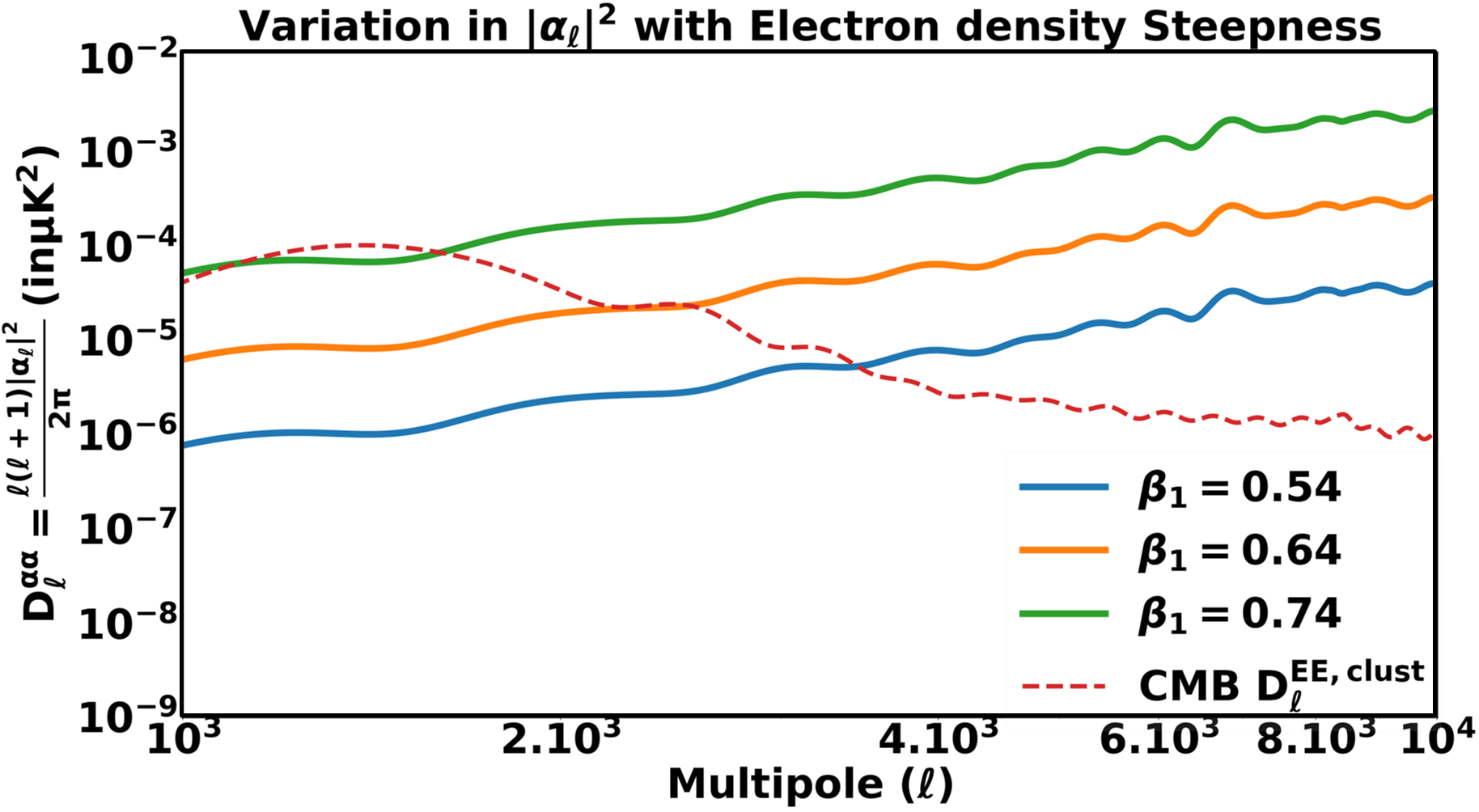}
         \caption{Variation in $\mathrm{|\alpha_{\ell}|^2}$ with Electron density steepening}
         \label{fig:b1_alp_vary}
     \end{subfigure}
     \begin{subfigure}[h]{0.45\textwidth}
         \centering
    \includegraphics[height=5cm,width=7cm]         {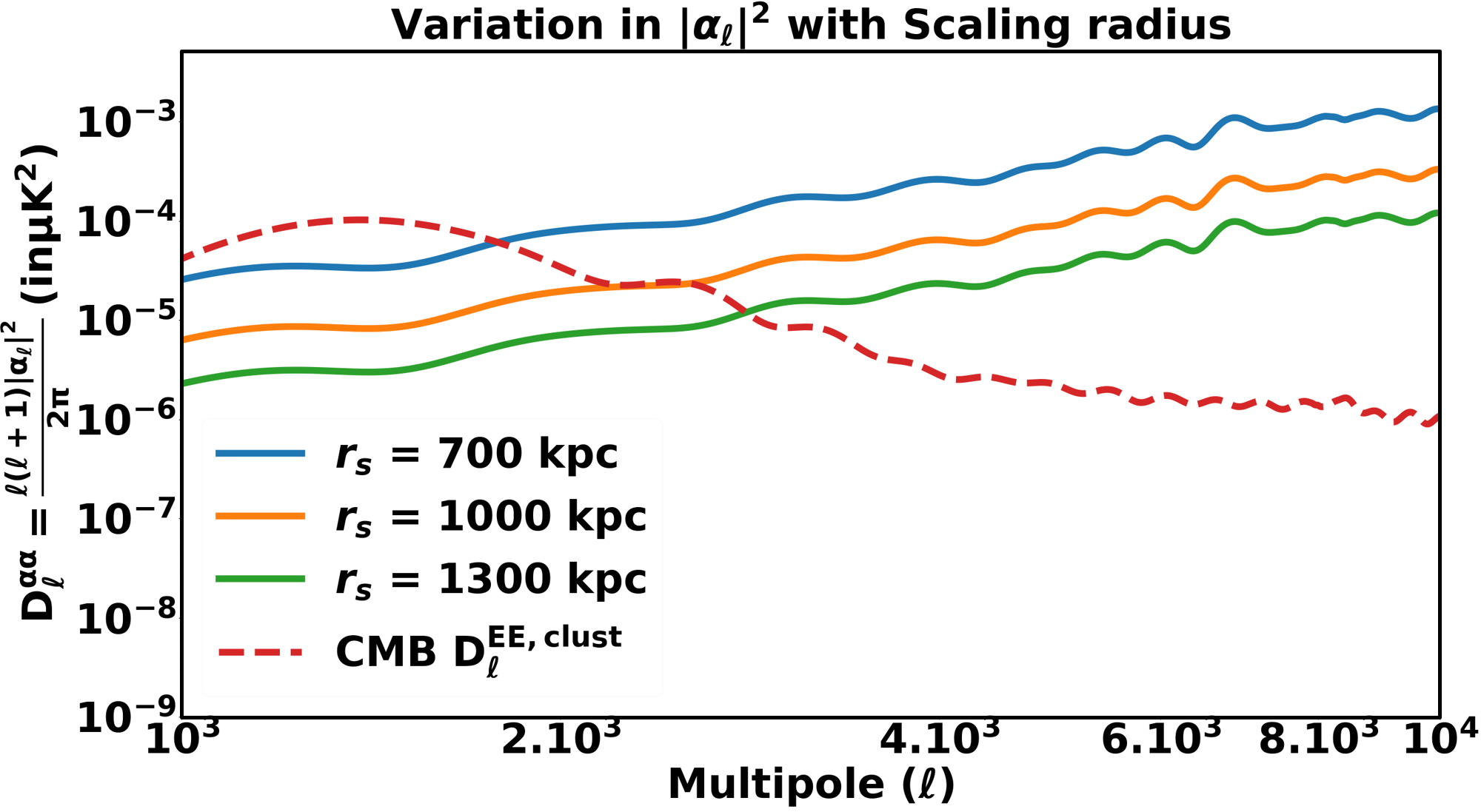}
         \caption{Variation in $\mathrm{|\alpha_{\ell}|^2}$ with Scaling radius}
         \label{fig:rs_alp_vary}
     \end{subfigure}
     \hspace{0.01cm} 
     \begin{subfigure}[h]{0.45\textwidth}
         \centering
    \includegraphics[height=5cm,width=7cm]{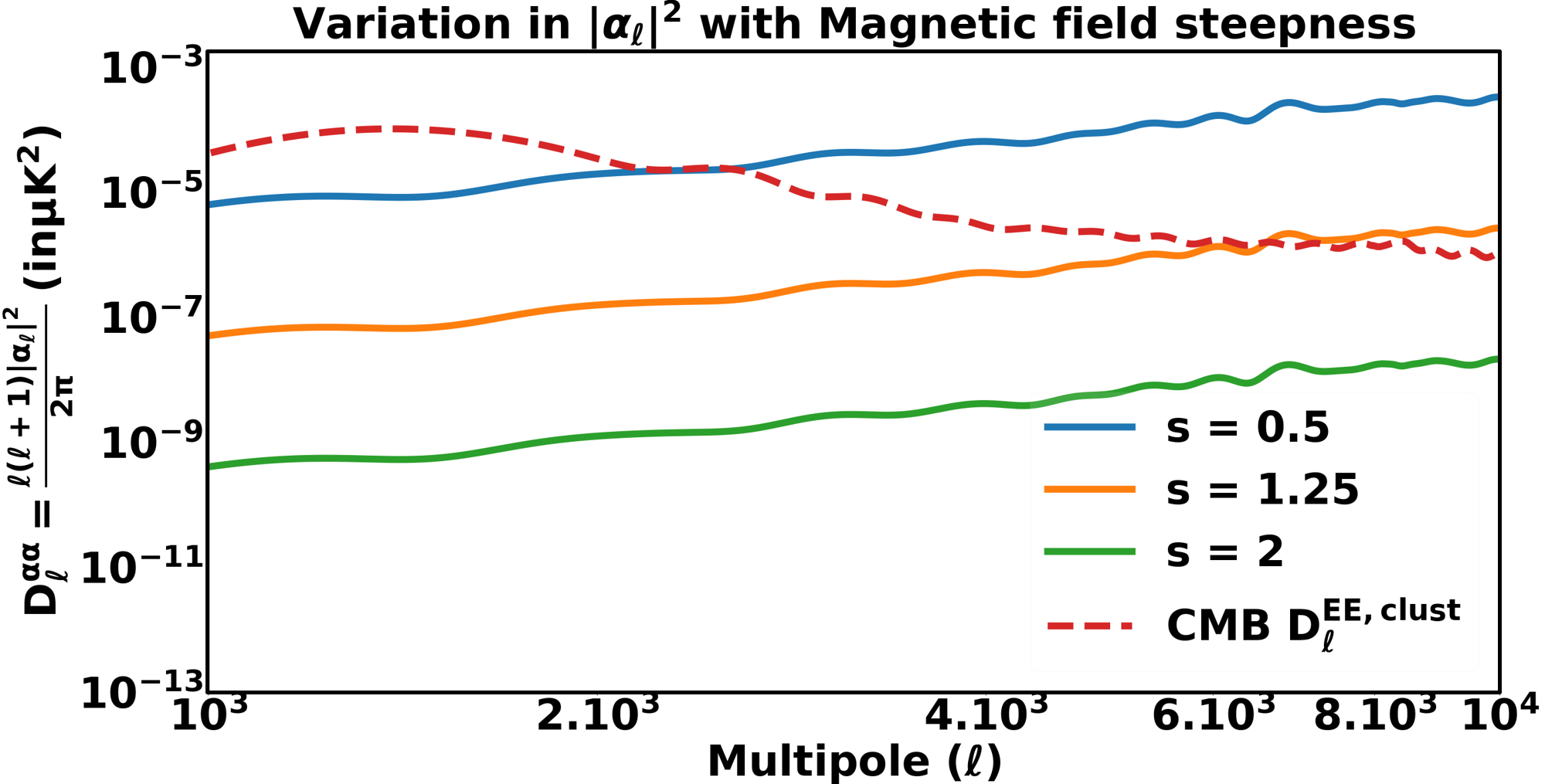}
         \caption{Variation in $\mathrm{|\alpha_{\ell}|^2}$ with Magnetic field steepening}
         \label{fig:s_alp_vary}
     \end{subfigure}
\caption{Astrophysical variation in $\mathrm{|\alpha_{\ell}|^2}$ due to  electron density (\ref{fig:n0_alp_vary}, \ref{fig:b1_alp_vary}, \ref{fig:rs_alp_vary}) and magnetic field  \ref{fig:s_alp_vary} profiles. The CMB polarization power spectrum at low angular scales (around the cluster region) is
also plotted.}
\label{fig: cluster depend}
\end{figure}

The features of the ALP power spectrum for a single cluster $|\alpha_{\ell}|^2$ depend on the host cluster and its properties that characterize the ALP distortion signal. 
The variation of the ALP power spectrum from a single cluster $|\alpha_{\ell}|^2$ with the photon-ALP coupling constant and the frequency of observation is shown in Fig. \ref{fig: alpha_f_g}.  The ALP distortion signal scales as  $g^2_{a\gamma}$ and $\nu I_{\mathrm{cmb}}(\nu)$ (Eq. \ref{eq:gamma_ad}) for frequencies of observation in the microwave and radio bands. Thus, the power spectrum $|\alpha_{\ell}|^2$ varies as $g^4_{a\gamma}$ and $\nu ^2 I_{\mathrm{cmb}}(\nu)^2$, where $I_{\mathrm{cmb}}(\nu)$ is the Planck black-body function (see Eq.\ref{eq:Distort}). Also shown in Fig.\ref{fig: alpha_f_g} is the CMB power spectrum at small scales around the cluster region.

The probability of conversion at a location is inversely proportional to the gradient of electron density. An increase in the electron density ({$n_e \propto n_0$) (Fig. \ref{fig:n0_alp_vary}}) reduces the ALP power spectrum strength. The electron density  steepness parameter($\beta_1$) and scaling radius ($r_{s}$) control the  rate at which the electron density varies with distance from the cluster center. The spectrum increases at high multipoles with decreasing scaling radius and increasing steepness parameter as the electron density reduces in the outer regions of the cluster (Fig. \ref{fig:b1_alp_vary} and \ref{fig:rs_alp_vary}). 
Also, the spectrum is directly proportional to the square of the magnetic field. So, a steep decrease in magnetic field (parameterized by "$s$") leads to a weaker spectrum as the magnetic field reduces in the outer regions of the cluster where the conversion probability is high (Fig. \ref{fig:s_alp_vary}).  

The effect of change in profiles on $|\alpha_{\ell}|^2$ is dominated by the effect in the production of low mass ALPs. This is because the strength of the ALP distortion signal is generally high for low mass ALPs, owing to higher conversion probabilities in the outer regions of the cluster with low electron densities ($\propto |\nabla n_e(r)|^{-1}$). This, in turn, shows up in the effect on $|\alpha_{\ell}|^2$.
If ALPs of masses in only a sub-range of what we have considered ($10^{-15} - 10^{-11}$ eV) are formed, the dependence of the ALP background spectrum on these profile parameters can change, especially in the case of formation of only high mass ALPs. This is considered in a separate work \cite{mehta2024power}. 

The power spectrum $|\alpha_{\ell}|^2$ also depends on the redshift of the host cluster, scaling proportionally to $1 + z$, where $z$ is the redshift of the cluster. This happens as the conversion depends on the photon frequency at the conversion location (which is within the cluster) and this photon then gets cosmologically redshifted on its travel from the cluster to us due to Hubble expansion \cite{Dodelson:2003ft}.

The shape of $|\alpha_{\ell}|^2$'s at high multipoles depends on the magnetic field orientation that characterizes the polarization of the ALP signal. If the magnetic field orientation is uniformly random, as in our analysis, $|\alpha_{\ell}|^2$ scales as $\ell^0$ ($D_{\ell}$ as $\ell^2$) at high multipoles.
\begin{figure}[h!]
     \centering
\includegraphics[height=6.5cm,width=12.5cm]{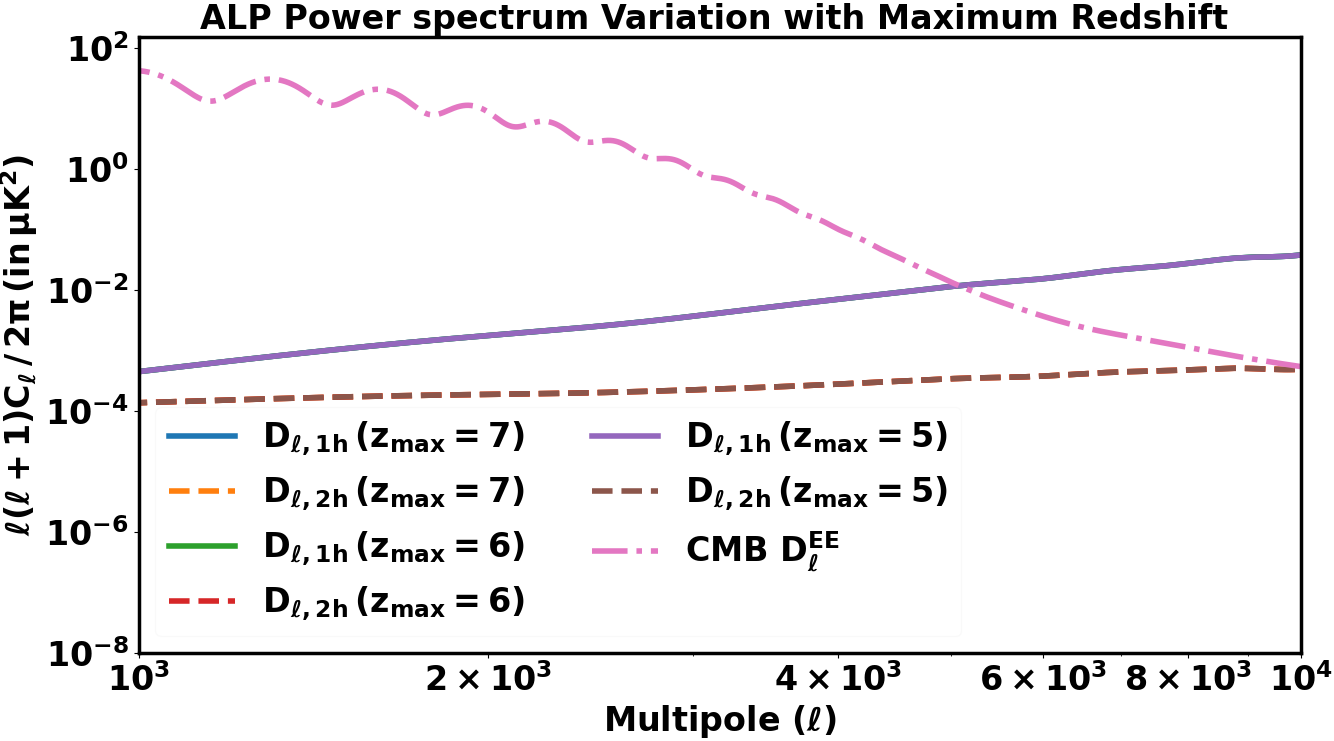}
    \caption{Variation of the ALP power spectrum (solid: one-halo, dashed: two-halo) for various maximum redshifts. The plots are overlapping due to nearly the same strength of the signal with a change in the maximum redshift. This points towards the low contribution to the power spectrum from high redshifts. The CMB E-mode spectrum is also plotted for reference.}
       \label{fig:zmax_halo_vary}
\end{figure}
\section{ALP background spectrum Variation with maximum redshift}\label{sec:z_vary} 
In principle, clusters from very high redshifts will contribute to the diffused spectrum. But this contribution will be very low and the power spectrum is nearly independent of the maximum redshift  $z_{\mathrm{max}}$ up to which we consider the distribution of clusters as can be seen in Fig. \ref{fig:zmax_halo_vary} as the spectra at various redshifts overlap. The halo mass function decreases significantly with increasing redshift and hence, the number of clusters contributing at very high redshifts ($z_{\mathrm{max}} > 3.5$) to the one and two-halo terms decreases significantly. Also, as a polarized photon travels from high redshifts, it may get depolarized due to multiple scatterings in galaxy clusters. Hence, it is mainly the polarized distortion signals from low redshift clusters that remain intact and contribute to the background ALP spectrum. This explains the choice of  $z_{\mathrm{max}}$ of 7 for our analysis as it reduces computation time. 

\section{Power spectrum estimation from a CMB map} \label{sec:estim_deri}
We follow the derivation given in \cite{Dodelson:2003ft,hu2002cosmic}. The fluctuations in the CMB or in any signal (foregrounds, ALP distortion, SZ effect, etc.) can be expanded in terms of spherical harmonics as:
\begin{equation}\label{eq:sphdec}
    \Delta^{\mathrm{net}} = \sum_{\ell = 0}^{\infty} \sum_{m=-\ell}^{\ell} a_{\ell m}Y_{\ell m}(\theta,\phi).
\end{equation} 
The power spectrum is given as:
\begin{equation}\label{eq:cl_alm}
\langle a_{\ell m}^{\mathrm{net*}} a_{\ell' m'}^{\mathrm{net}} \rangle= C_{\ell}^{\mathrm{net}}\delta_{\ell \ell'} \delta_{m m'} .
\end{equation}
Here we consider the  CMB primordial fluctuations and fluctuations from ALP conversion and foregrounds: 
\begin{equation}\label{eq:pertparts}
    \Delta^{\mathrm{net}} = \Delta^{\mathrm{cmb}} + \Delta^{\mathrm{ax}} + \Delta^{\mathrm{fg}}.
\end{equation}
The net power spectrum is given as the ensemble average (with $i$ and $j$ running over the components):
\begin{equation}\label{eq:clnet}
C_{\ell}^{\mathrm{net}} = \langle a_{\ell m}^{\mathrm{ax*}} a_{\ell m}^{\mathrm{ax}} \rangle = \sum_{j} \sum_{i} \langle a_{\ell m}^{i*} a_{\ell m}^{j} \rangle = \langle a_{\ell m}^{\mathrm{cmb*}} a_{\ell m}^{\mathrm{cmb}} \rangle + \langle a_{\ell m}^{\mathrm{ax*}} a_{\ell m}^{\mathrm{ax}} \rangle + \langle a_{\ell m}^{\mathrm{fg*}} a_{\ell m}^{\mathrm{fg}} \rangle.
\end{equation}
Here we have considered the signals to be independent of each other and neglected the cross terms.

The finite beam resolution and instrumental noise of the experiment change the spherical harmonic  
coefficient as:
\begin{equation}\label{eq:alm_obs}
  a_{\ell m}^{\mathrm{obs}} = B_{\ell}(a_{\ell m}^{\mathrm{cmb}} + a_{\ell m}^{\mathrm{ax}} + a_{\ell m}^{\mathrm{fg}}) + \eta_{\ell m} ,  
\end{equation}

where $B_{\ell} = \exp(-\ell(\ell+1)\theta_{\mathrm{beam}}^2 / 2)$ and $\eta_{\ell m}$ are the fourier coefficients introduced due to instrumental noise.

The coefficients $a_{\ell m}^i$'s are assumed to follow a Gaussian distribution with mean zero and variance given by the corresponding $C_{\ell}^{i}$'s, i.e.,

\begin{equation}\label{eq:prob_alm_cl}
    P(a_{\ell m}^{i}|C_{\ell} ^{i} ) = \frac{1}{\sqrt{2\pi C_{\ell}^i}} \exp \left( - \frac{|a_{\ell m}^{i}|^2}{2 C_{\ell}^i} \right).
\end{equation}

The noise power spectrum is obtained as:
\begin{equation}\label{eq:Nl}
\langle \eta_{\ell m}* \eta_{\ell 'm'} \rangle = N_{\ell}^{i} \delta_{\ell \ell '} \delta_{mm'}.    
\end{equation}

The $a_{\ell m}^{\mathrm{obs}}$s are assumed to follow a  Gaussian distribution with mean $B_{\ell} \sum_{i}a_{\ell m}^i$ and variance $N_{\ell}$ given as:
\begin{equation}\label{eq:prov_almobs_alm}
P(a_{{\ell}m}^{\mathrm{obs}}|\{a_{\ell m}^i \} ) = \frac{1}{\sqrt{2\pi N_{\ell}}} \exp \left( - \frac{|a_{\ell m}^{\mathrm{obs}} - \sum_{i} B_{\ell} a_{\ell m}^i|^2}{2 N_{\ell}} \right).    
\end{equation}

Using Baye's theorem, we have:
\begin{equation}\label{eq:estimlike}
\begin{split}
    P(a_{\ell m}^{\mathrm{obs}}|\{C_{\ell}^i \} ) = \prod_{m=-\ell}^{\ell} \int\int\int \prod_{i} \mathrm{d} \, a_{\ell m}^{i}  P(a_{\ell m}^{\mathrm{obs}} |  a_{ \ell m}^i) P(a_{\ell m}^{i}|C_{\ell}^{i} ), 
    \\
     = [2\pi (B_{\ell}^2 \sum_{i}C_{\ell}^{i} + N_{\ell})]^{-(2\ell + 1)/2} \exp \left[ \sum_{m = -\ell}^{\ell}  -\frac{|a_{\ell m}^{\mathrm{obs}}|^2}{2(\sum_{i}C_{\ell}^{i} B_{\ell}^2 + N_{\ell})} \right] .   
\end{split}
\end{equation}

We find the maximum likelihood estimator for the power spectrum of the component $i$ by differentiating with respect to $C_{\ell}^{i}$ and setting equal to zero. It can be written as:
\begin{equation}\label{eq:estimator}
  \tilde{C_{\ell}^{i}} = B_{\ell}^{-2} \left[ \frac{1}{2\ell + 1} \sum_{m = -\ell}^{\ell} |a_{\ell m}^{\mathrm{obs}}|^2 - N_{\ell} \right] - \sum_{j \neq i} C_{\ell}^{j}.  
\end{equation}

\section{Constraints on ALP coupling constant using template matching of foregrounds}\label{sec:tempmatch}
\begin{figure}[h!]
     \centering
\includegraphics[height=6cm,width=11cm]{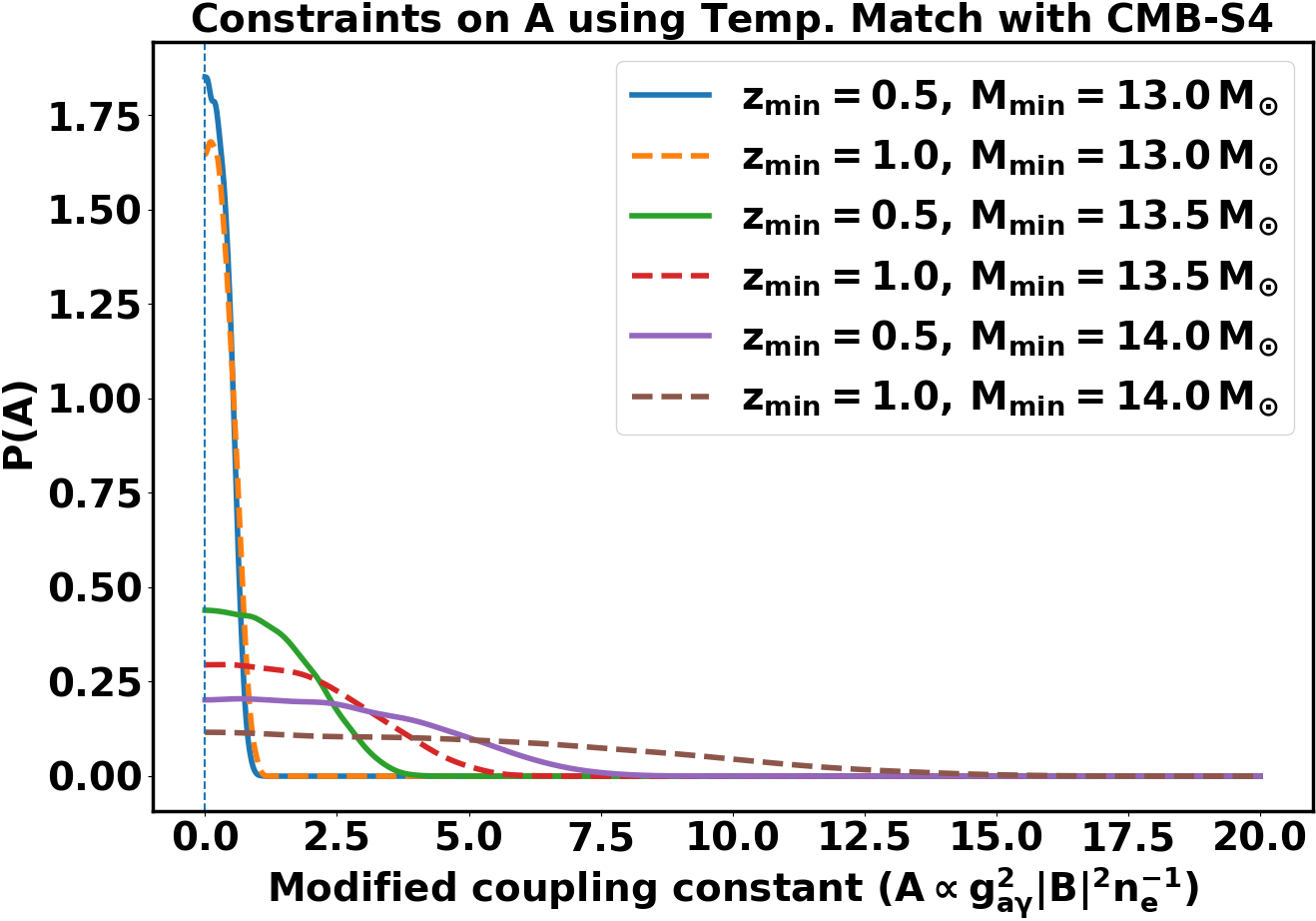}
    \caption{Constraints from CMB-S4 using Template Matching at 145 GHz for $z_{\mathrm{min}} = 0.5$ and $z_{\mathrm{min}} = 1$}
       \label{fig:s4_temp}
\end{figure}
Template matching assumes the scaling of the foregrounds power spectra with frequencies and can be used to obtain stronger bounds on the modified coupling constant using a lower beam size. We perform template matching to look for the improvement in constraints.
\begin{figure}[h!]
     \centering
\includegraphics[height=6cm,width=11cm]{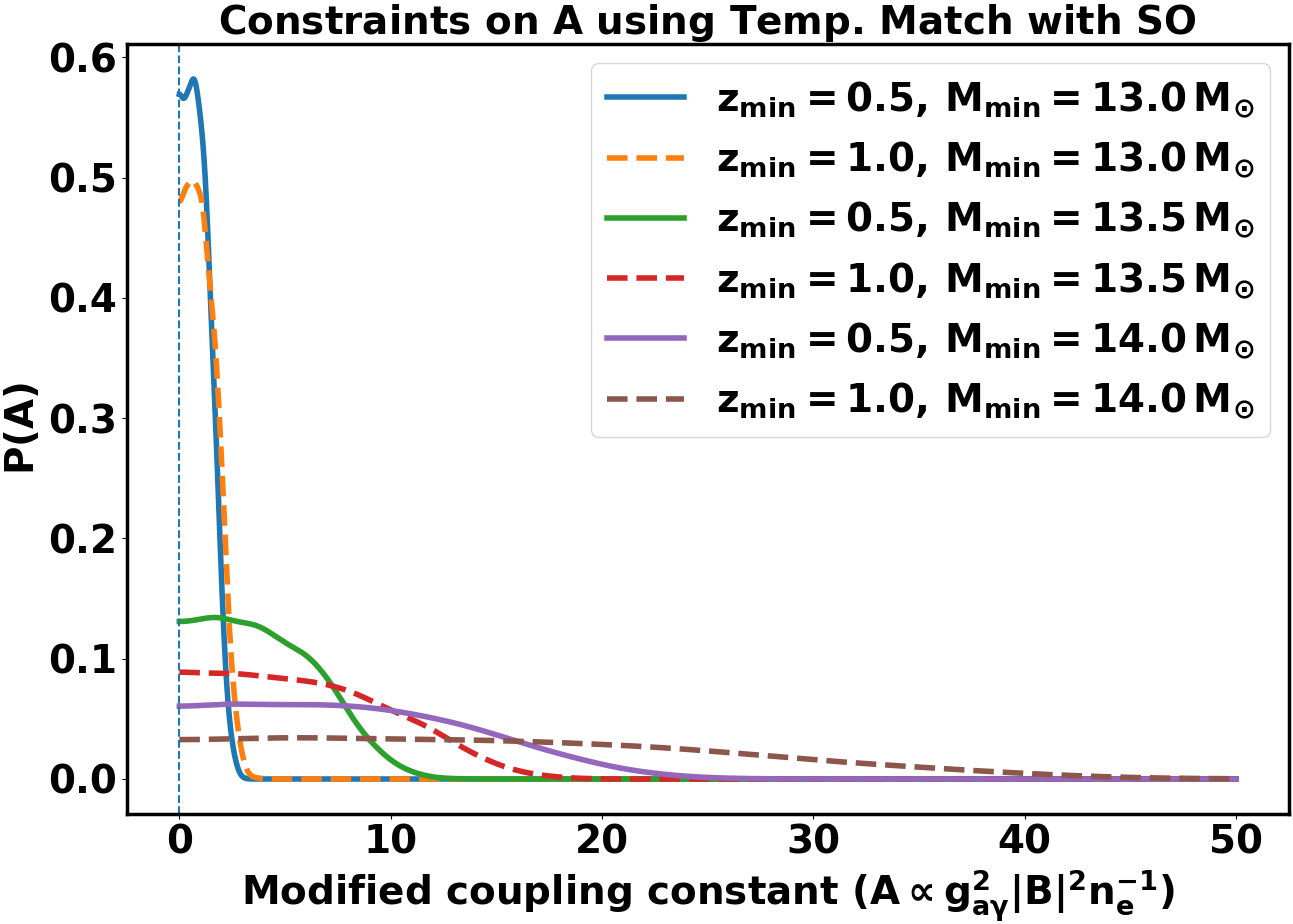}
    \caption{Constraints from SO using Template Matching at 145 GHz for $z_{\mathrm{min}} = 0.5$ and $z_{\mathrm{min}} = 1$}
       \label{fig:so_temp}
\end{figure}
Even after masking, the effect of these foregrounds remains even at high latitudes. Thus, a modelling of these foregrounds is required to account for their impact on the power spectrum. These are modelled using high frequencies ($\nu > 200$ GHz) for dust, while low frequencies ($\nu < 70$ GHz) for synchrotron emission. 

\begin{figure}[h!]
     \centering
\includegraphics[height=6cm,width=11cm]{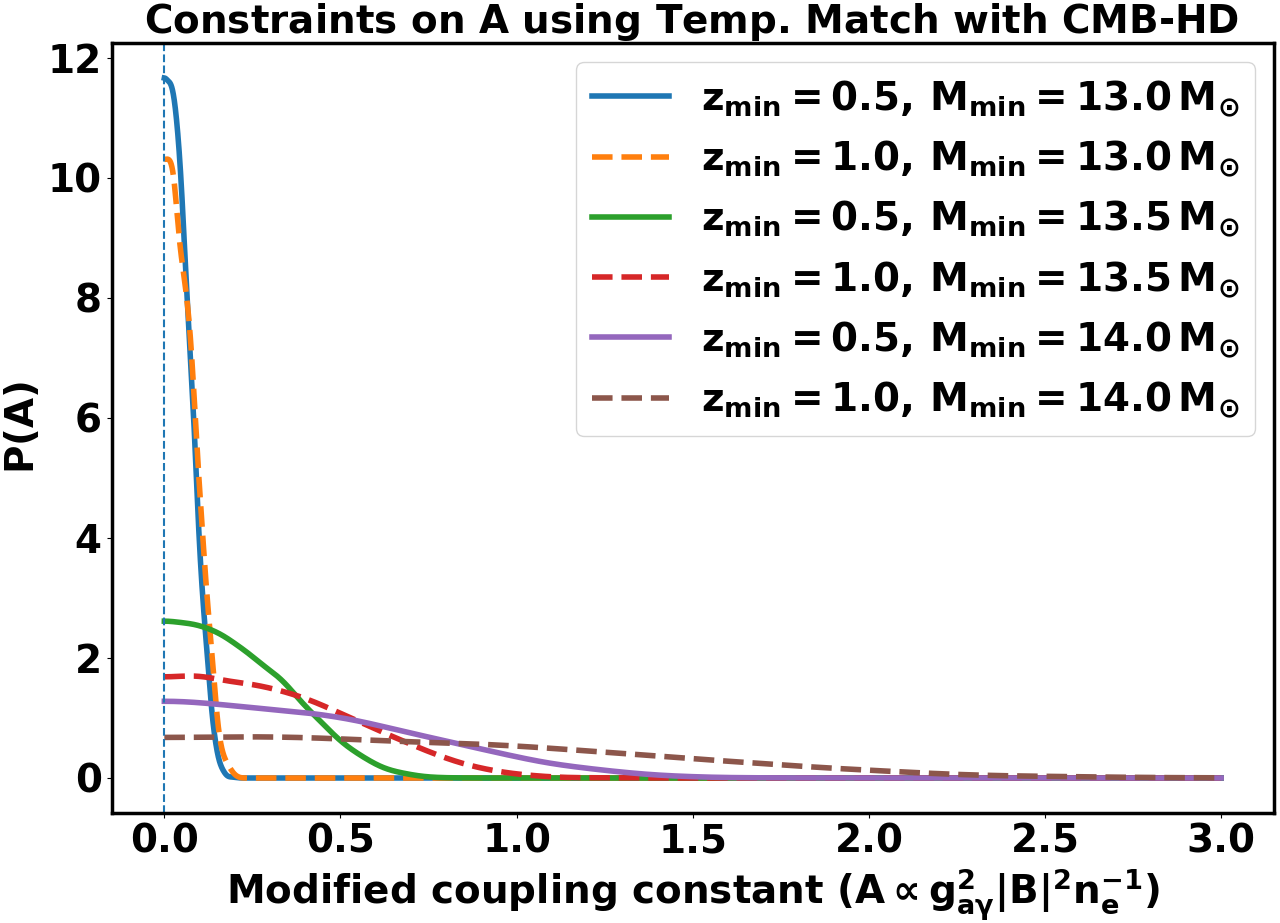}
    \caption{Constraints from CMB-HD using Template Matching at 150 GHz for $z_{\mathrm{min}} = 0.5$ and $z_{\mathrm{min}} = 1$}
       \label{fig:hd_temp}
\end{figure}

We perform a template matching of foregrounds for CMB-S4 configuration to look for any improvement in the bounds on modified coupling constant. We assume we know the scaling of galactic synchrotron and dust emissions with frequencies in the microwave and radio regions of the EM spectra. To model the "s-3" synchrotron model used in our mock data, we use the following equation to account for its curved index \cite{Thorne_2017}:
\begin{equation}
   C_{\ell}^{\mathrm{syn}}(\nu) = A_{\ell} \left( \frac{\nu}{\nu_0} \right)^{2\beta_s + 2C \ln (\nu / \nu_c)}.
    \label{eq:s3}
\end{equation}

The curved spectral index shows a steepening or flattening with frequency above the frequency $\nu_c$. The fiducial values used are $C = -0.052$, $\nu_c = 23$ GHz and $\beta_s = -3$. 

The modified black-body function with $\beta_d = 1.58$ is used to fit the dust model "d-3":
\begin{equation}
C_{\ell}^{\mathrm{dust}}(\nu) = A_{\ell}\nu ^{2\beta_d} B^2_{\nu}(T).
   \label{eq:d3}
\end{equation}

By modelling the foreground emissions at low (for synchrotron) and high (for dust) frequencies, the scaling with frequency can be used to obtain the contribution of synchrotron and dust at the frequencies 90 - 160 GHz. This method assumes the scaling of foregrounds with frequencies for multipole range $\ell > \ell_{\mathrm{max}}$ corresponding to the beam of the instrument at the matched frequency of 145 GHz for CMB-S4 and SO, while 150 GHz for CMB-HD. 

We find the fiducial (non-ALPs) contribution of CMB and foregrounds at the required frequency by making different map realizations without ALP signal and calculating the mean power spectrum of those maps.
This enables us to scale the ALP diffused spectrum at the matched frequency with respect to the residual of the mock data spectrum and fiducial spectrum. This scaling is compared against the covariance at that frequency (Eq.\ref{eq:covar}) to obtain bounds on the modified ALP coupling constant. 

The constraints obtained on the modified ALP coupling constant $A$ are shown for various CMB surveys for minimum redshifts $z_{\mathrm{min}} = 0.5$ and $z_{\mathrm{min}} = 1$ in Figs. \ref{fig:s4_temp} (CMB-S4: 
$A < 0.654$ with $z_{\mathrm{min}} = 0.5$ and $A < 0.734$ with $z_{\mathrm{min}} = 1$ ), \ref{fig:so_temp} (SO: $A < 2.005$ with $z_{\mathrm{min}} = 0.5$ and $A < 2.325$ with $z_{\mathrm{min}} = 1$ ) and \ref{fig:hd_temp} (CMB-HD: $A < 0.115$ with $z_{\mathrm{min}} = 0.5$ and $A < 0.131$ with $z_{\mathrm{min}} = 1$ ). The constraints are expected with CMB-HD giving the tightest constraints. The constraints from template matching are stronger, as compared to ILC for all three detectors (SO, CMB-S4, and CMB-HD).

\begin{figure}[h!]
     \centering
\includegraphics[height=7cm,width=11cm]{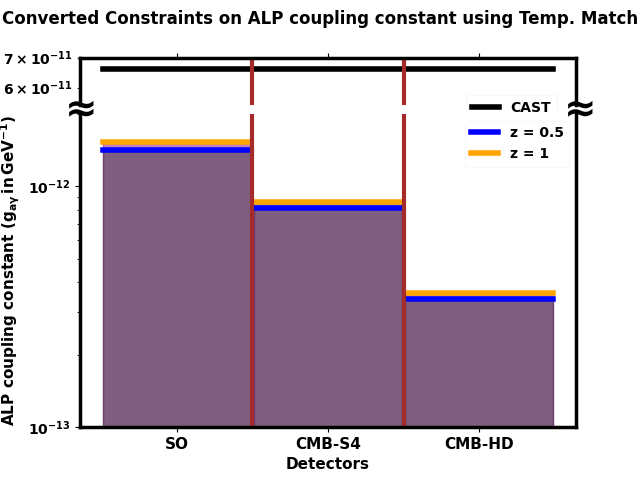}
    \caption{Converted ALP coupling constant constraints for various detectors using Template Matching for $z_{\mathrm{min}} = 0.5$ and $z_{\mathrm{min}} = 1$}
       \label{fig:ga_cons_tmp}
\end{figure}

The converted constraints (95\% C.I.) on ALP coupling constant $g_{a\gamma}$ using template matching are shown in Fig.\ref{fig:ga_cons_tmp}. With template matching, we will be able to get the following bounds with $z_{\mathrm{min}} = 0.5$:  SO -  $g_{a\gamma} < 1.416 \times 10^{-12} \, \mathrm{GeV^{-1}}$; CMB-S4 -  
$g_{a\gamma} < 8.087 \times 10^{-13} \, \mathrm{GeV^{-1}}$; CMB-HD - $g_{a\gamma} < 3.391 \times 10^{-13} \, \mathrm{GeV^{-1}}$, while the bounds with  $z_{\mathrm{min}} = 1$ go as: 
$$\mathrm{SO} :  g_{a\gamma} < 1.525 \times 10^{-12} \mathrm{GeV^{-1}} ;$$ 
$$\mathrm{CMB-S4} :  
g_{a\gamma} < 8.567 \times 10^{-13} \, \mathrm{GeV^{-1}};$$ 
$$\mathrm{CMB-HD} : g_{a\gamma} < 3.619 \times 10^{-13} \, \mathrm{GeV^{-1}}.$$

Both 
template matching and ILC assume the shape of the foregrounds power spectrum for $\ell > \ell_{\mathrm{max}}$, but since the beam size is larger for ILC (lower $\ell_{\mathrm{max}}$), template matching provides better bounds with a higher $\ell_{\mathrm{max}}$ corresponding to a lower beam size at the matched frequency for all detectors, but is less reliable as the modelling of foregrounds at different frequencies may not be precise leading to bias in the constraints.

\acknowledgments
    This work is a part of the $\langle \texttt{data|theory}\rangle$ \texttt{Universe-Lab}, supported by the TIFR  and the Department of Atomic Energy, Government of India. The authors express their gratitude to the TIFR CCHPC facility for meeting the computational needs. Furthermore, we  would also like to thank the Simons Observatory (SO), CMB-S4 and CMB-HD and collaborations for providing the instrument noise and beam resolutions. 
 Also, the following packages were used for this work: Astropy \cite{astropy:2013,astropy:2022,astropy:2018},
, NumPy \cite{harris2020array}
CAMB \cite{2011ascl.soft02026L}, SciPy \cite{2020SciPy-NMeth}, SymPy \cite{10.7717/peerj-cs.103}, Matplotlib \cite{Hunter:2007}, HEALPix (Hierarchical Equal Area isoLatitude Pixelation of a sphere)\footnote{Link to the HEALPix website http://healpix.sf.net}\cite{2005ApJ...622..759G,Zonca2019}, PySM \cite{Thorne_2017} and Cluster Toolkit \cite{2022ascl.soft09004M}.

\bibliographystyle{JHEP.bst}
\bibliography{references}

\providecommand{\href}[2]{#2}\begingroup\raggedright\begin{thebibliography}{10}

\bibitem{Dodelson:2003ft}
S.~Dodelson, \emph{{Modern Cosmology}}. Academic Press, Amsterdam, 2003.

\bibitem{Fixsen_2009}
D.~J. Fixsen, \emph{The temperature of the cosmic microwave background}, \href{https://doi.org/10.1088/0004-637x/707/2/916}{\emph{The Astrophysical Journal} {\bfseries 707} (2009) 916–920}.

\bibitem{PhysRevD.104.022003}
{\scshape SPT-3G Collaboration} collaboration, D.~Dutcher, L.~Balkenhol, P.~A.~R. Ade, Z.~Ahmed, E.~Anderes and A.~J. e.~a. Anderson, \emph{Measurements of the $e$-mode polarization and temperature-$e$-mode correlation of the cmb from spt-3g 2018 data}, \href{https://doi.org/10.1103/PhysRevD.104.022003}{\emph{Phys. Rev. D} {\bfseries 104} (2021) 022003}.

\bibitem{Ade_2019}
P.~Ade, J.~Aguirre and Z.~e.~a. Ahmed, \emph{The simons observatory: science goals and forecasts}, \href{https://doi.org/10.1088/1475-7516/2019/02/056}{\emph{Journal of Cosmology and Astroparticle Physics} {\bfseries 2019} (2019) 056–056}.

\bibitem{abazajian2016cmbs4}
K.~N. Abazajian, P.~Adshead, Z.~Ahmed and S.~W.~A. et~al., \emph{Cmb-s4 science book, first edition},  2016.

\bibitem{sehgal2019cmbhd}
N.~Sehgal, S.~Aiola, Y.~Akrami and K.~B. et~al., \emph{Cmb-hd: An ultra-deep, high-resolution millimeter-wave survey over half the sky},  2019.

\bibitem{cyr2024cmbspectraldistortionsmultimessenger}
B.~Cyr, \emph{Cmb spectral distortions: A multimessenger probe of the primordial universe},  2024.

\bibitem{Lucca:2023seh}
M.~Lucca, \emph{{The future of cosmology? A case for CMB spectral distortions}}, Ph.D. thesis, U. Brussels, 2023.
\newblock \href{https://arxiv.org/abs/2307.08513}{{\ttfamily 2307.08513}}.

\bibitem{chluba2016spectral}
J.~Chluba, \emph{Which spectral distortions does $\lambda$cdm actually predict?}, {\emph{Monthly Notices of the Royal Astronomical Society} {\bfseries 460} (2016) 227}.

\bibitem{chluba2013distinguishing}
J.~Chluba, \emph{Distinguishing different scenarios of early energy release with spectral distortions of the cosmic microwave background}, {\emph{Monthly Notices of the Royal Astronomical Society} {\bfseries 436} (2013) 2232}.

\bibitem{mather1994measurement}
J.~C. Mather, E.~Cheng, D.~A. Cottingham, R.~Eplee~Jr, D.~J. Fixsen, T.~Hewagama et~al., \emph{Measurement of the cosmic microwave background spectrum by the cobe firas instrument}, {\emph{The Astrophysical Journal, Part 1 (ISSN 0004-637X), vol. 420, no. 2, p. 439-444} {\bfseries 420} (1994) 439}.

\bibitem{2014PTEP.2014fB107T}
H.~{Tashiro}, \emph{{CMB spectral distortions and energy release in the early universe}}, \href{https://doi.org/10.1093/ptep/ptu066}{\emph{Progress of Theoretical and Experimental Physics} {\bfseries 2014} (2014) 06B107}.

\bibitem{berezhiani1991cosmology}
Z.~Berezhiani and M.~Y. Khlopov, \emph{Cosmology of spontaneously broken gauge family symmetry with axion solution of strong cp-problem}, {\emph{Zeitschrift f{\"u}r Physik C Particles and Fields} {\bfseries 49} (1991) 73}.

\bibitem{dine1983not}
M.~Dine and W.~Fischler, \emph{The not-so-harmless axion}, {\emph{Physics Letters B} {\bfseries 120} (1983) 137}.

\bibitem{abbott1983cosmological}
L.~F. Abbott and P.~Sikivie, \emph{A cosmological bound on the invisible axion}, {\emph{Physics Letters B} {\bfseries 120} (1983) 133}.

\bibitem{preskill1983cosmology}
J.~Preskill, M.~B. Wise and F.~Wilczek, \emph{Cosmology of the invisible axion}, {\emph{Physics Letters B} {\bfseries 120} (1983) 127}.

\bibitem{Ghosh:2023xhs}
A.~Ghosh and P.~Konar, \emph{{Precision prediction at the LHC of a democratic up-family philic KSVZ axion model}},  \href{https://arxiv.org/abs/2305.08662}{{\ttfamily 2305.08662}}.

\bibitem{marsh2014model}
D.~J. Marsh and J.~Silk, \emph{A model for halo formation with axion mixed dark matter}, {\emph{Monthly Notices of the Royal Astronomical Society} {\bfseries 437} (2014) 2652}.

\bibitem{marsh2017axions}
D.~J. Marsh, \emph{Axions and alps: a very short introduction}, {\emph{arXiv preprint arXiv:1712.03018} (2017) }.

\bibitem{tashiro2013constraints}
H.~Tashiro, J.~Silk and D.~J. Marsh, \emph{Constraints on primordial magnetic fields from cmb distortions in the axiverse}, {\emph{Physical Review D} {\bfseries 88} (2013) 125024}.

\bibitem{Ghosh:2022rta}
A.~Ghosh, P.~Konar and R.~Roshan, \emph{{Top-philic dark matter in a hybrid KSVZ axion framework}}, \href{https://doi.org/10.1007/JHEP12(2022)167}{\emph{JHEP} {\bfseries 12} (2022) 167} [\href{https://arxiv.org/abs/2207.00487}{{\ttfamily 2207.00487}}].

\bibitem{1992SvJNP..55.1063B}
Z.~G. {Berezhiani}, A.~S. {Sakharov} and M.~Y. {Khlopov}, \emph{{Primordial background of cosmological axions.}}, {\emph{Soviet Journal of Nuclear Physics} {\bfseries 55} (1992) 1063}.

\bibitem{khlopov1999nonlinear}
M.~Y. Khlopov, A.~Sakharov and D.~Sokoloff, \emph{The nonlinear modulation of the density distribution in standard axionic cdm and its cosmological impact}, {\emph{Nuclear Physics B-Proceedings Supplements} {\bfseries 72} (1999) 105}.

\bibitem{sakharov1994nonhomogeneity}
A.~Sakharov and M.~Y. Khlopov, \emph{The nonhomogeneity problem for the primordial axion field}, {\emph{Physics of Atomic Nuclei} {\bfseries 57} (1994) 485}.

\bibitem{sakharov1996large}
A.~Sakharov, D.~Sokoloff and M.~Y. Khlopov, \emph{Large-scale modulation of the distribution of coherent oscillations of a primordial axion field in the universe}, {\emph{Physics of Atomic Nuclei} {\bfseries 59} (1996) }.

\bibitem{Mukherjee_2019}
S.~Mukherjee, R.~Khatri and B.~D. Wandelt, \emph{Constraints on non-resonant photon-axion conversion from the planck satellite data}, \href{https://doi.org/10.1088/1475-7516/2019/06/031}{\emph{Journal of Cosmology and Astroparticle Physics} {\bfseries 2019} (2019) 031–031}.

\bibitem{Mukherjee_2018}
S.~Mukherjee, R.~Khatri and B.~D. Wandelt, \emph{Polarized anisotropic spectral distortions of the cmb: galactic and extragalactic constraints on photon-axion conversion}, \href{https://doi.org/10.1088/1475-7516/2018/04/045}{\emph{Journal of Cosmology and Astroparticle Physics} {\bfseries 2018} (2018) 045–045}.

\bibitem{Mukherjee_2020}
S.~Mukherjee, D.~N. Spergel, R.~Khatri and B.~D. Wandelt, \emph{A new probe of axion-like particles: Cmb polarization distortions due to cluster magnetic fields}, \href{https://doi.org/10.1088/1475-7516/2020/02/032}{\emph{Journal of Cosmology and Astroparticle Physics} {\bfseries 2020} (2020) 032–032}.

\bibitem{osti_22525054}
M.~Schlederer and G.~Sigl, \emph{Constraining alp-photon coupling using galaxy clusters}, \href{https://doi.org/10.1088/1475-7516/2016/01/038}{\emph{Journal of Cosmology and Astroparticle Physics} {\bfseries 2016} (2016) }.

\bibitem{song2024polarizationsignalsaxionphotonresonant}
N.~Song, L.~Su and L.~Wu, \emph{Polarization signals from axion-photon resonant conversion in neutron star magnetosphere},  2024.

\bibitem{GOVONI_2004}
F.~GOVONI and L.~FERETTI, \emph{Magnetic fields in clusters of galaxies}, \href{https://doi.org/10.1142/s0218271804005080}{\emph{International Journal of Modern Physics D} {\bfseries 13} (2004) 1549–1594}.

\bibitem{Birkinshaw_1999}
M.~Birkinshaw, \emph{The sunyaev–zel’dovich effect}, \href{https://doi.org/10.1016/s0370-1573(98)00080-5}{\emph{Physics Reports} {\bfseries 310} (1999) 97–195}.

\bibitem{2014ApJS..210....9B}
M.~{Bilicki}, T.~H. {Jarrett}, J.~A. {Peacock}, M.~E. {Cluver} and L.~{Steward}, \emph{{Two Micron All Sky Survey Photometric Redshift Catalog: A Comprehensive Three-dimensional Census of the Whole Sky}}, \href{https://doi.org/10.1088/0067-0049/210/1/9}{\emph{\apjs} {\bfseries 210} (2014) 9} [\href{https://arxiv.org/abs/1311.5246}{{\ttfamily 1311.5246}}].

\bibitem{mehta2024power}
H.~Mehta and S.~Mukherjee, \emph{A power spectrum approach to search for axion-like particles from resolved galaxy clusters using cmb as a backlight}, {\emph{arXiv preprint arXiv:2405.08878} (2024) }.

\bibitem{Mehta:2024:new3}
H.~Mehta and S.~Mukherjee, \emph{Spectrax: A new multi-band data analysis framework to search for axion-like particles using galaxy clusters}, {\emph{(under preparation)} (2024) }.

\bibitem{COORAY_2002}
A.~COORAY and R.~SHETH, \emph{Halo models of large scale structure}, \href{https://doi.org/10.1016/s0370-1573(02)00276-4}{\emph{Physics Reports} {\bfseries 372} (2002) 1–129}.

\bibitem{Eriksen_2004}
H.~K. Eriksen, A.~J. Banday, K.~M. Gorski and P.~B. Lilje, \emph{On foreground removal from thewilkinson microwave anisotropy probedata by an internal linear combination method: Limitations and implications}, \href{https://doi.org/10.1086/422807}{\emph{The Astrophysical Journal} {\bfseries 612} }.

\bibitem{ilc2008internal}
R.~Vio and P.~Andreani, \emph{"internal linear combination" method for the separation of cmb from galactic foregrounds in the harmonic domain},  2008.

\bibitem{2016}
P.~A.~R. Ade, N.~Aghanim, M.~Arnaud, M.~Ashdown and J.~e.~a. Aumont, \emph{Planck2015 results: Xiii. cosmological parameters}, \href{https://doi.org/10.1051/0004-6361/201525830}{\emph{Astronomy \&; Astrophysics} {\bfseries 594} (2016) A13}.

\bibitem{1986rpa..book.....R}
G.~B. {Rybicki} and A.~P. {Lightman}, \emph{{Radiative Processes in Astrophysics}}. 1986.

\bibitem{2012MNRAS.419.1294C}
J.~{Chluba} and R.~A. {Sunyaev}, \emph{{The evolution of CMB spectral distortions in the early Universe}}, \href{https://doi.org/10.1111/j.1365-2966.2011.19786.x}{\emph{\mnras} {\bfseries 419} (2012) 1294} [\href{https://arxiv.org/abs/1109.6552}{{\ttfamily 1109.6552}}].

\bibitem{Hu_1997}
W.~Hu and M.~White, \emph{A cmb polarization primer}, \href{https://doi.org/10.1016/s1384-1076(97)00022-5}{\emph{New Astronomy} {\bfseries 2} (1997) 323–344}.

\bibitem{Smith_2007}
K.~M. Smith, O.~Zahn and O.~Doré, \emph{Detection of gravitational lensing in the cosmic microwave background}, \href{https://doi.org/10.1103/physrevd.76.043510}{\emph{Physical Review D} {\bfseries 76} (2007) }.

\bibitem{Raffelt:1996wa}
G.~G. Raffelt, \emph{{Stars as laboratories for fundamental physics}: {The astrophysics of neutrinos, axions, and other weakly interacting particles}}. 5, 1996.

\bibitem{mead2015accurate}
A.~Mead, J.~Peacock, C.~Heymans, S.~Joudaki and A.~Heavens, \emph{An accurate halo model for fitting non-linear cosmological power spectra and baryonic feedback models}, {\emph{Monthly Notices of the Royal Astronomical Society} {\bfseries 454} (2015) 1958}.

\bibitem{2011ascl.soft02026L}
A.~{Lewis} and A.~{Challinor}, ``{CAMB: Code for Anisotropies in the Microwave Background}.'' Astrophysics Source Code Library, record ascl:1102.026, Feb., 2011.

\bibitem{Tinker_2008}
J.~Tinker, A.~V. Kravtsov, A.~Klypin, K.~Abazajian, M.~Warren, G.~Yepes et~al., \emph{Toward a halo mass function for precision cosmology: The limits of universality}, \href{https://doi.org/10.1086/591439}{\emph{The Astrophysical Journal} {\bfseries 688} (2008) 709–728}.

\bibitem{Komatsu_1999}
E.~Komatsu and T.~Kitayama, \emph{Sunyaev-zeldovich fluctuations from spatial correlations between clusters of galaxies}, \href{https://doi.org/10.1086/312364}{\emph{The Astrophysical Journal} {\bfseries 526} (1999) L1–L4}.

\bibitem{Remazeilles_2021}
M.~Remazeilles, A.~Rotti and J.~Chluba, \emph{Peeling off foregrounds with the constrained moment ilc method to unveil primordial cmb b modes}, \href{https://doi.org/10.1093/mnras/stab648}{\emph{Monthly Notices of the Royal Astronomical Society} {\bfseries 503} (2021) 2478–2498}.

\bibitem{vacher2023high}
L.~Vacher, J.~Chluba, J.~Aumont, A.~Rotti and L.~Montier, \emph{High precision modeling of polarized signals: Moment expansion method generalized to spin-2 fields}, {\emph{Astronomy \& Astrophysics} {\bfseries 669} (2023) A5}.

\bibitem{vacher2022moment}
L.~Vacher, J.~Aumont, L.~Montier, S.~Azzoni, F.~Boulanger and M.~Remazeilles, \emph{Moment expansion of polarized dust sed: A new path towards capturing the cmb b-modes with litebird}, {\emph{Astronomy \& Astrophysics} {\bfseries 660} (2022) A111}.

\bibitem{carlstrom2002cosmology}
J.~E. Carlstrom, G.~P. Holder and E.~D. Reese, \emph{Cosmology with the sunyaev-zel’dovich effect}, {\emph{Annual Review of Astronomy and Astrophysics} {\bfseries 40} (2002) 643}.

\bibitem{Stompor_1999}
R.~Stompor and G.~Efstathiou, \emph{Gravitational lensing of cosmic microwave background anisotropies and cosmological parameter estimation}, \href{https://doi.org/10.1046/j.1365-8711.1999.02174.x}{\emph{Monthly Notices of the Royal Astronomical Society} {\bfseries 302} (1999) 735–747}.

\bibitem{aghanim2020planck}
N.~Aghanim, Y.~Akrami, M.~Ashdown, J.~Aumont, C.~Baccigalupi, M.~Ballardini et~al., \emph{Planck 2018 results-viii. gravitational lensing}, {\emph{Astronomy \& Astrophysics} {\bfseries 641} (2020) A8}.

\bibitem{Thorne_2017}
B.~Thorne, J.~Dunkley, D.~Alonso and S.~Næss, \emph{The python sky model: software for simulating the galactic microwave sky}, \href{https://doi.org/10.1093/mnras/stx949}{\emph{Monthly Notices of the Royal Astronomical Society} {\bfseries 469} (2017) 2821–2833}.

\bibitem{Remazeilles_2010}
M.~Remazeilles, J.~Delabrouille and J.-F. Cardoso, \emph{Cmb and sz effect separation with constrained internal linear combinations: Cmb and sz separation with constrained ilc}, \href{https://doi.org/10.1111/j.1365-2966.2010.17624.x}{\emph{Monthly Notices of the Royal Astronomical Society} {\bfseries 410} (2010) 2481–2487}.

\bibitem{zhang2024constrained}
Z.~Zhang, Y.~Liu, S.-Y. Li, H.~Li and H.~Li, \emph{A constrained nilc method for cmb b mode observations}, {\emph{Journal of Cosmology and Astroparticle Physics} {\bfseries 2024} (2024) 014}.

\bibitem{kogut2011primordial}
A.~Kogut, D.~Fixsen, D.~Chuss, J.~Dotson, E.~Dwek, M.~Halpern et~al., \emph{The primordial inflation explorer (pixie): a nulling polarimeter for cosmic microwave background observations}, {\emph{Journal of Cosmology and Astroparticle Physics} {\bfseries 2011} (2011) 025}.

\bibitem{kogut2016primordial}
A.~Kogut, J.~Chluba, D.~J. Fixsen, S.~Meyer and D.~Spergel, \emph{The primordial inflation explorer (pixie)},  in \emph{Space telescopes and instrumentation 2016: optical, infrared, and millimeter wave}, vol.~9904, pp.~331--353, SPIE, 2016.

\bibitem{kogut2024primordial}
A.~Kogut, E.~Switzer, D.~Fixsen, N.~Aghanim, J.~Chluba, D.~Chuss et~al., \emph{The primordial inflation explorer (pixie): Mission design and science goals}, {\emph{arXiv preprint arXiv:2405.20403} (2024) }.

\bibitem{mondino2024axion}
C.~Mondino, D.~P{\^\i}rvu, J.~Huang and M.~C. Johnson, \emph{Axion-induced patchy screening of the cosmic microwave background}, {\emph{arXiv preprint arXiv:2405.08059} (2024) }.

\bibitem{Komatsu2000CMBAF}
E.~Komatsu, T.~Kitayama, A.~R{\'e}fr{\'e}gier, D.~N. Spergel and U.~li~Pen, \emph{Cmb anisotropy from spatial correlations of clusters of galaxies}, {\emph{arXiv: Astrophysics} (2000) 2189}.

\bibitem{Tremmel_2017}
M.~Tremmel, M.~Karcher, F.~Governato, M.~Volonteri, T.~R. Quinn, A.~Pontzen et~al., \emph{The romulus cosmological simulations: a physical approach to the formation, dynamics and accretion models of smbhs}, \href{https://doi.org/10.1093/mnras/stx1160}{\emph{Monthly Notices of the Royal Astronomical Society} {\bfseries 470} (2017) 1121–1139}.

\bibitem{Dav__2019}
R.~Davé, D.~Anglés-Alcázar, D.~Narayanan, Q.~Li, M.~H. Rafieferantsoa and S.~Appleby, \emph{simba: Cosmological simulations with black hole growth and feedback}, \href{https://doi.org/10.1093/mnras/stz937}{\emph{Monthly Notices of the Royal Astronomical Society} {\bfseries 486} (2019) 2827–2849}.

\bibitem{Vikhlinin_2006}
A.~Vikhlinin, A.~Kravtsov, W.~Forman, C.~Jones, M.~Markevitch, S.~S. Murray et~al., \emph{Chandrasample of nearby relaxed galaxy clusters: Mass, gas fraction, and mass‐temperature relation}, \href{https://doi.org/10.1086/500288}{\emph{The Astrophysical Journal} {\bfseries 640} (2006) 691–709}.

\bibitem{mcdonald2013growth}
M.~McDonald, B.~Benson, A.~Vikhlinin, B.~Stalder, L.~Bleem, T.~De~Haan et~al., \emph{The growth of cool cores and evolution of cooling properties in a sample of 83 galaxy clusters at 0.3< z< 1.2 selected from the spt-sz survey}, {\emph{The Astrophysical Journal} {\bfseries 774} (2013) 23}.

\bibitem{Carilli_2004}
C.~Carilli and S.~Rawlings, \emph{Motivation, key science projects, standards and assumptions}, \href{https://doi.org/10.1016/j.newar.2004.09.001}{\emph{New Astronomy Reviews} {\bfseries 48} (2004) 979–984}.

\bibitem{bonafede2010galaxy}
A.~Bonafede, L.~Feretti, M.~Murgia, F.~Govoni, G.~Giovannini and V.~Vacca, \emph{Galaxy cluster magnetic fields from radio polarized emission}, {\emph{arXiv preprint arXiv:1009.1233} (2010) }.

\bibitem{bohringer2016cosmic}
H.~B{\"o}hringer, G.~Chon and P.~P. Kronberg, \emph{The cosmic large-scale structure in x-rays (classix) cluster survey-i. probing galaxy cluster magnetic fields with line of sight rotation measures}, {\emph{Astronomy \& Astrophysics} {\bfseries 596} (2016) A22}.

\bibitem{hu2002cosmic}
W.~Hu and S.~Dodelson, \emph{Cosmic microwave background anisotropies}, {\emph{Annual Review of Astronomy and Astrophysics} {\bfseries 40} (2002) 171}.

\bibitem{astropy:2013}
{Astropy Collaboration}, T.~P. {Robitaille}, E.~J. {Tollerud}, P.~{Greenfield}, M.~{Droettboom} and E.~e.~a. {Bray}, \emph{{Astropy: A community Python package for astronomy}}, \href{https://doi.org/10.1051/0004-6361/201322068}{\emph{\aap} {\bfseries 558} (2013) A33} [\href{https://arxiv.org/abs/1307.6212}{{\ttfamily 1307.6212}}].

\bibitem{astropy:2022}
{Astropy Collaboration}, A.~M. {Price-Whelan}, P.~L. {Lim} and N.~e.~a. {Earl}, \emph{{The Astropy Project: Sustaining and Growing a Community-oriented Open-source Project and the Latest Major Release (v5.0) of the Core Package}}, \href{https://doi.org/10.3847/1538-4357/ac7c74}{\emph{\apj} {\bfseries 935} (2022) 167} [\href{https://arxiv.org/abs/2206.14220}{{\ttfamily 2206.14220}}].

\bibitem{astropy:2018}
{Astropy Collaboration}, A.~M. {Price-Whelan}, B.~M. {Sip{\H{o}}cz}, H.~M. {G{\"u}nther}, P.~L. {Lim} and S.~M. e.~a. {Crawford}, \emph{{The Astropy Project: Building an Open-science Project and Status of the v2.0 Core Package}}, \href{https://doi.org/10.3847/1538-3881/aabc4f}{\emph{\aj} {\bfseries 156} (2018) 123} [\href{https://arxiv.org/abs/1801.02634}{{\ttfamily 1801.02634}}].

\bibitem{harris2020array}
C.~R. Harris, K.~J. Millman, S.~J. van~der Walt, R.~Gommers, P.~Virtanen and D.~C. et~al., \emph{Array programming with {NumPy}}, \href{https://doi.org/10.1038/s41586-020-2649-2}{\emph{Nature} {\bfseries 585} (2020) 357}.

\bibitem{2020SciPy-NMeth}
P.~Virtanen, R.~Gommers, T.~E. Oliphant, M.~Haberland, T.~Reddy and D.~e.~a. Cournapeau, \emph{{{SciPy} 1.0: Fundamental Algorithms for Scientific Computing in Python}}, \href{https://doi.org/10.1038/s41592-019-0686-2}{\emph{Nature Methods} {\bfseries 17} (2020) 261}.

\bibitem{10.7717/peerj-cs.103}
A.~Meurer, C.~P. Smith and M.~e.~a. Paprocki, \emph{Sympy: symbolic computing in python}, \href{https://doi.org/10.7717/peerj-cs.103}{\emph{PeerJ Computer Science} {\bfseries 3} (2017) e103}.

\bibitem{Hunter:2007}
J.~D. Hunter, \emph{Matplotlib: A 2d graphics environment}, \href{https://doi.org/10.1109/MCSE.2007.55}{\emph{Computing in Science \& Engineering} {\bfseries 9} (2007) 90}.

\bibitem{2005ApJ...622..759G}
K.~M. {G{\'o}rski}, E.~{Hivon}, A.~J. {Banday}, B.~D. {Wandelt}, F.~K. {Hansen}, M.~{Reinecke} et~al., \emph{{HEALPix: A Framework for High-Resolution Discretization and Fast Analysis of Data Distributed on the Sphere}}, \href{https://doi.org/10.1086/427976}{\emph{\apj} {\bfseries 622} (2005) 759} [\href{https://arxiv.org/abs/arXiv:astro-ph/0409513}{{\ttfamily arXiv:astro-ph/0409513}}].

\bibitem{Zonca2019}
A.~Zonca, L.~Singer, D.~Lenz, M.~Reinecke, C.~Rosset, E.~Hivon et~al., \emph{healpy: equal area pixelization and spherical harmonics transforms for data on the sphere in python}, \href{https://doi.org/10.21105/joss.01298}{\emph{Journal of Open Source Software} {\bfseries 4} (2019) 1298}.

\bibitem{2022ascl.soft09004M}
T.~{McClintock}, ``{Cluster Toolkit: Tools for analyzing galaxy clusters}.'' Astrophysics Source Code Library, record ascl:2209.004, Sept., 2022.

\end{thebibliography}\endgroup

\end{document}